\documentclass[journal,onecolumn]{IEEEtran}
\usepackage{booktabs}
\usepackage{amsmath}
\usepackage{amsmath}                
\usepackage{mathtools} 
\usepackage{amsthm}                 
\usepackage{amssymb}
\usepackage{thmtools}
\usepackage{kpfonts}
\usepackage[geometry]{ifsym}
\usepackage{dsfont}
\usepackage{multirow}
\usepackage[mathscr]{euscript}
\usepackage{graphicx}
\usepackage{color}
\usepackage[inline]{enumitem}
\usepackage{framed}
\usepackage{float}
\usepackage{longtable}
\usepackage{cite}
\usepackage{enumitem}
\usepackage{caption} 
\usepackage{subcaption} 
\usepackage{xcolor}

\usepackage[colorlinks=true, citecolor=blue, linkcolor=blue]{hyperref}       
\usepackage[font=itshape]{quoting}
\usepackage{lscape}
\usepackage{makecell}
\usepackage[section]{placeins}

\usepackage{graphicx}
\usepackage
{listings}
\usepackage{comment}
\usepackage{framed} 
\usepackage{nomencl} 
\makenomenclature

\declaretheoremstyle[%
  headfont=\bfseries,%
  headpunct={:},%
  notefont=\normalfont\bfseries,%
  notebraces={--~}{},
    qed=$\blacksquare
$,
]{definitionstyle}
\theoremstyle{definition}
\declaretheorem[style=definitionstyle,name=Definition]{defn}

\theoremstyle{definition}

\theoremstyle{plain}
\theoremstyle{remark}

\begin{document}

\title{Spatio-Temporal Life Cycle Analysis of Electrolytic H$_2$ Production in Australia under Time-Varying CO$_2$ Management Schemes}

\author{Niraj Gohil, Alexander Franke, Nawshad Haque, Amro M. Farid}
\maketitle

\begin{abstract}
The transition to sustainable energy is critical for addressing global climate change. Hydrogen production, particularly via electrolysis, has emerged as a key solution, offering the potential for low-carbon energy across various sectors. This paper presents a novel approach to enhancing hydrogen production by aligning it with periods of low-carbon intensity on the electricity grid. Leveraging real-time data from the Electricity Mapping database and real-time electricity cost data from the AEMO database, the model dynamically adjusts hydrogen output to minimize both emissions and production costs. Furthermore, the integration of hydrogen tax credits significantly enhances cost-effectiveness, offering a viable pathway for widespread adoption. A comprehensive Life Cycle Assessment (LCA) framework is employed to assess the environmental impacts, emphasizing the need for real-time data incorporation to more accurately reflect hydrogen production's carbon footprint. The study concludes that dynamic, real-time operation, coupled with financial incentives, provides a promising method to enhance the sustainability and economic viability of hydrogen production.
\end{abstract}
\par\null\par

\section{Introduction}
As the global energy demand continues to rise, the urgency to combat climate change through clean and sustainable energy solutions becomes more pressing. Among the most promising solutions is hydrogen production, particularly via electrolysis, due to its potential to serve as a low-carbon energy carrier across sectors such as transportation, industrial processes, and power generation. Given the critical need to reduce carbon emissions, hydrogen is poised to play a pivotal role in the global energy transition.
However, traditional hydrogen production methods are often inefficient in terms of both emissions and costs. This paper proposes a novel solution that enhances hydrogen production through real-time data-driven decision-making, aligning production with periods of low-carbon intensity electricity. By integrating real-time carbon intensity data from the Electricity Mapping database\cite{electricitymaps_datasets} and real-time electricity cost data from the Australian Energy Market Operator (AEMO)\cite{aemo_nem_data_dashboard}, the model dynamically adjusts hydrogen output based on both environmental and economic factors. This dynamic operation not only minimizes emissions but also reduces the cost of hydrogen production.  Moreover, the inclusion of hydrogen tax credits significantly improves the cost-effectiveness of green hydrogen, making it more competitive against traditional fossil fuel-based alternatives. To assess the full environmental and economic implications, a comprehensive Life Cycle Assessment (LCA) is conducted, enhanced using real-time data to evaluate hydrogen’s sustainability more accurately.  This study demonstrates that by leveraging real-time operation and financial incentives, hydrogen production can achieve substantial improvements in both environmental performance and cost-effectiveness, offering a scalable solution for the transition to a sustainable energy future.

\subsection{Literature Gap:}
The transition to green hydrogen production is critical for achieving global sustainability goals and reducing the carbon footprint across various sectors. Electrolysis, a leading technology for producing hydrogen, has been widely studied, but challenges remain in optimizing it for environmental and economic performance. This review examines the literature on hydrogen production via electrolysis, Life Cycle Assessment (LCA), real-time optimization, and policy incentives such as hydrogen tax credits, all of which form the foundation for the approach developed in this paper.  Hydrogen production via electrolysis, particularly when powered by renewable energy, has been recognized as a promising pathway for decarbonization. According to Bhandari et al. (2014)\cite{Bhandari:2014:00}, electrolysis has the potential to deliver zero-emission hydrogen if it operates using low-carbon electricity. However, one of the key challenges remains the variability in the carbon intensity of grid electricity. Ursua et al. (2012)\cite{Ursua:2012:00} highlight that the carbon footprint of hydrogen produced through electrolysis depends heavily on the energy mix of the grid at any given time, pointing to the need for dynamic operation based on real-time carbon intensity data.

Life Cycle Assessment (LCA) is a critical tool used to evaluate the environmental impact of hydrogen production throughout its lifecycle Goren et al. (2025) \cite{Goren:2025:00}. Traditional LCAs rely on static, annual average data, which fail to account for fluctuations in carbon intensity throughout the day. Franke and Farid (2024) \cite{franke:2024:00} address this gap by incorporating time-varying carbon intensities into the LCA for hydrogen generation, demonstrating that hydrogen production aligned with periods of low-carbon electricity can significantly reduce emissions.

The economic viability of hydrogen production is closely tied to its operational costs, which are largely driven by electricity prices. Frank et al. (2022) \cite{FRANK:2021:00} highlight that real-time electricity cost data can be integrated into hydrogen production models to optimize for both cost and emissions. By dynamically adjusting hydrogen production based on periods of lower electricity prices, operators can reduce the overall cost of production. The role of real-time data in enhancing both emissions and costs is further emphasized by Zhiyuan et al. (2019) \cite{Zhiyuan:2024:00}, who demonstrate that dynamic operation of hydrogen production can lead to substantial cost savings, especially when aligned with low-carbon energy availability.

Policy mechanisms, such as hydrogen tax credits, are essential for driving the adoption of green hydrogen technologies. Reports from IRENA (2021)\cite{irena:2021:00} and DOE (2022)\cite{doe:2022:00} emphasize the importance of financial incentives in reducing the cost of hydrogen production and making it competitive with traditional, fossil-fuel-based energy. Hydrogen tax credits can provide substantial cost savings, making green hydrogen more attractive for large-scale industrial applications. The Hydrogen Production Tax Incentive Consultation Paper (2024)\cite{treasury2024hydrogen:2024:00} offers a practical example of how tax credits can support the economic viability of hydrogen production, especially when combined with real-time operation strategies that reduce both cost and emissions.

The integration of hydrogen tax credits with real-time data-driven production models creates a synergy that can significantly enhance the cost-effectiveness of green hydrogen. However, the literature primarily focuses on either emissions or cost reduction in isolation. Few studies explore how real-time data integration can maximize the impact of financial incentives like tax credits. By incorporating these two elements, this paper seeks to address this gap and present a more comprehensive approach to optimizing hydrogen production.  This review highlights the critical gaps in the current literature: the reliance on static data in traditional LCAs, the limited integration of real-time carbon intensity data, and the need for synergistic use of policy incentives and technological optimization. This paper builds on these gaps by presenting a model that dynamically enhances hydrogen production using real-time carbon intensity and electricity cost data while integrating hydrogen tax credits to enhance both environmental and economic performance.

\subsection{Original Contribution}
This paper makes several significant contributions to the field of hydrogen production and sustainability optimization. 
\begin{enumerate}
\item \textbf{Real-Time Carbon Intensity Alignment:} The paper introduces a novel methodology that aligns hydrogen production with periods of low-carbon electricity, utilizing real-time data from the Electricity Mapping database \cite{electricitymaps_datasets}. This dynamic approach reduces the CO$_2$ emissions and operational inefficiencies associated with hydrogen production.  
\item \textbf{Incorporation of Real-Time Price Data:} Unlike conventional studies, this research integrates real-time electricity price data from the AEMO database\cite{aemo_nem_data_dashboard}. By synchronizing hydrogen production with periods of lower electricity prices, the model achieves substantial cost reductions, enhancing both emissions and production cost.
\item \textbf{Policy Incentives for Cost Effectiveness:} The study integrates the economic benefits of hydrogen tax credits. This financial strategy enhances the competitiveness of green hydrogen, making it a more feasible alternative to carbon-intensive energy sources.
\item \textbf{Dynamic Life Cycle Assessment (LCA):} While previous studies often rely on static LCA data, this paper incorporates real-time data into the LCA framework. This approach provides a more accurate and comprehensive assessment of the environmental impact of hydrogen production, offering valuable insights for future clean energy projects.
\item \textbf{Pathway for Large-Scale Adoption:}  By presenting a model that integrates technological, environmental, and economic factors, the paper offers a scalable solution that addresses both the challenges of hydrogen production and its role in the broader energy transition. This approach facilitates the widespread adoption of hydrogen as a sustainable energy source, thereby contributing to global efforts to achieve carbon neutrality.
\end{enumerate}

\subsection{Paper Outline}
The remainder of the paper is organized as follows.  Sec. \ref{Sec:Background} provides background knowledge of how to conduct a spatio-temporal life cycle analysis using Model-Based Systems Engineering and Hetero-functional Graph Theory.  Sec. \ref{Sec:LCA-AUS} demonstrates this spatio-temporal life cycle analysis method on electrolytic hydrogen production in Australia.  Sec. \ref{Sec:CaseStudy} then extends this preliminary demonstration to a year-long life cycle analysis of electrolytic hydrogen production across five states in Australia:  New South Wales, Queensland, South Australia, Tasmania, and Victoria.  Sec. \ref{Sec:Conclusion} brings the work to a conclusion.  

\section{Background:  Life Cycle Analysis by MBSE and Hetero-functional Graph Theory}\label{Sec:Background}
To support the analytical discussion in the following sections, this section describes a recently developed method for conducting spatio-temporal life cycle analysis according to Niraj et al. (2025) \cite{Niraj.G:2025:00}. It uses Model-Based Systems Engineering Block Definition Diagrams in Sec. \ref{Sec:BackgroundMBSEBDD}, Activity Diagrams in Sec. \ref{Sec:BackgroundMBSEACT}, and Hetero-functional Graph Theory in Sec. \ref{Sec:BackgroundHFGT} database.  

\subsection{Model-Based Systems Engineering for Life Cycle Analysis: 
 Block Definition Diagram}\label{Sec:BackgroundMBSEBDD}
While a complete introduction to Model-Based Systems Engineering (MBSE) and SysML is beyond the scope of this paper, the essential elements are introduced to understand how they may be applied to life cycle analysis.  More specifically, the Block Definition Diagram and the Activity Diagram are discussed below.  

\begin{figure}[H]
\centering
\includegraphics[width=1\linewidth]{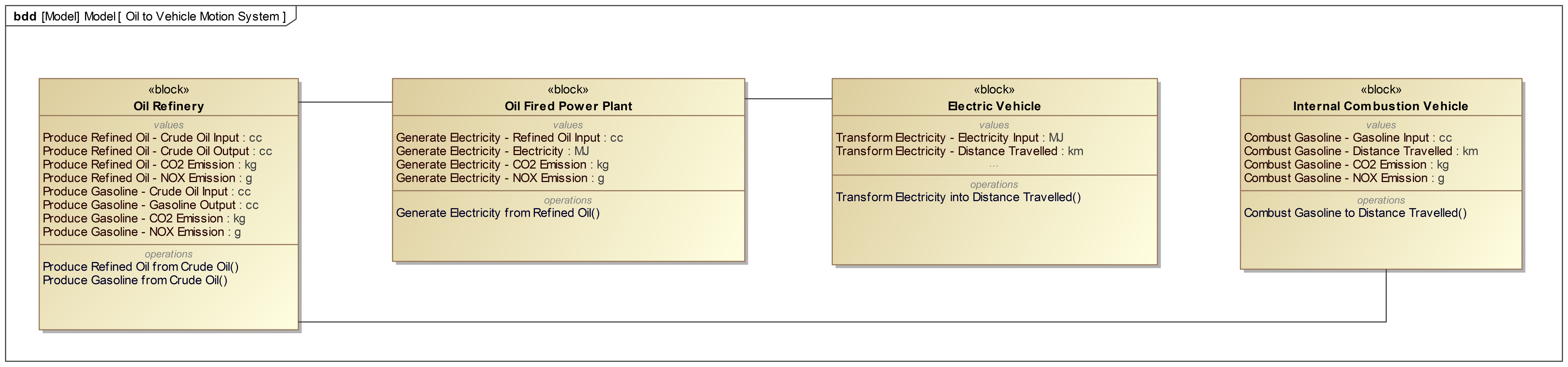}
\caption{A block definition diagram illustrating the Oil to Vehicle Motion System model, including its parts and data attributes.}
\label{Fig:BDDtoOilVehicleMotion}
\end{figure}

A Block Definition Diagram (BDD) is used to model a system’s form or its parts.  An example BDD is shown in Fig. \ref{Fig:BDDtoOilVehicleMotion}.  In the context of this paper, a BDD contains (at a minimum): 
\begin{itemize}
\item \textbf{Blocks:} that represent system elements, components, or subsystems.
\item \textbf{Attributes:} that represent characteristics or properties of each block.
\item \textbf{Operations:} that represent functions, activities, or processes that a block can perform.  Each combination of an operation in a block describes a capability.  
\end{itemize}
In the context of Fig. \ref{Fig:BDDtoOilVehicleMotion}, a BDD, entitled ``Oil to Vehicle Motion System," is used to capture the data required for the life cycle assessment of the distance traveled by an electric vehicle versus an internal combustion vehicle.  The system encompasses an oil refinery, an oil-fired power plant, an electric vehicle (EV), and an internal combustion engine (ICV). The diagram outlines each block's input and output parameters, including crude oil, refined oil, electricity, gasoline, emissions (CO$_2$ and NO$_x$), and vehicle range in kilometers, showcasing the interconnected pathways for energy conversion and emissions output.

\subsection{Model-Based Systems Engineering for Life Cycle Analysis: Activity Diagram}\label{Sec:BackgroundMBSEACT}
\begin{figure}[H]
\centering
\includegraphics[width=1\linewidth]{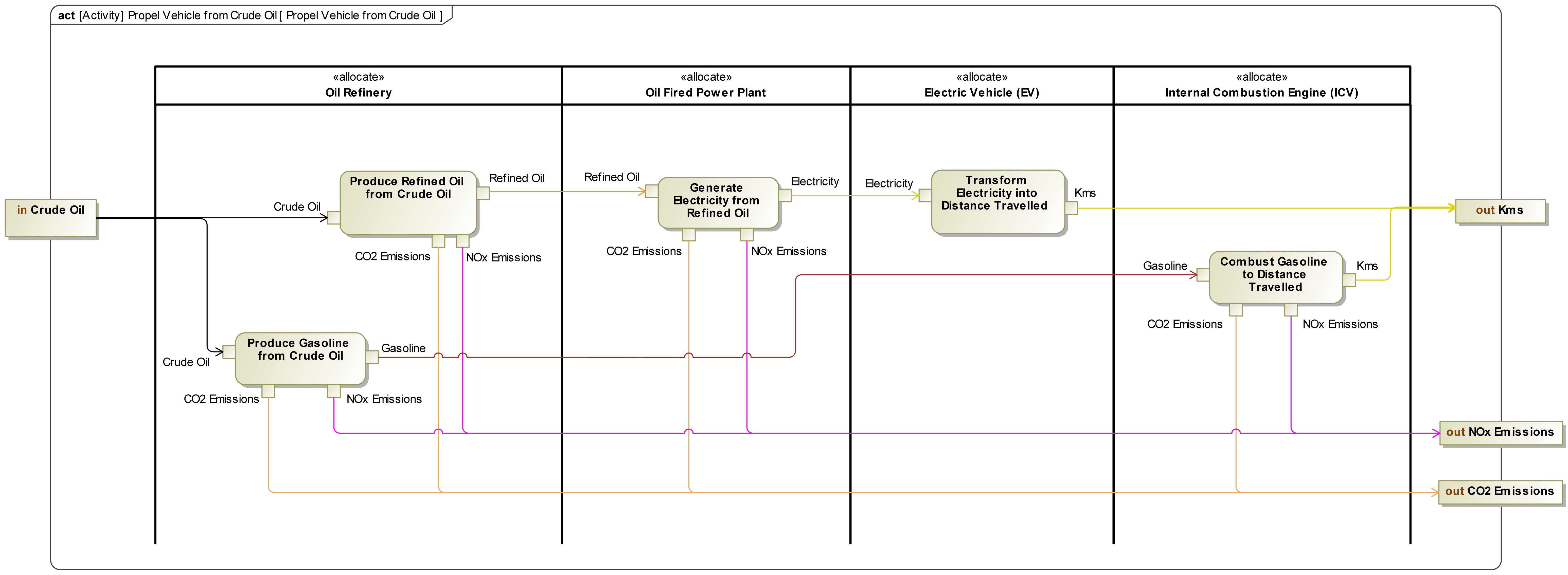}
\caption{An activity diagram with swimlanes illustrating the refining of oil, the production of gasoline, the generation of electricity, the transformation of electricity into motion, and the combustion of gasoline for motion.  Each swimlane represents a system component responsible for carrying out each action.  The activity diagram also shows crude oil as the input, and distance travelled,  CO$_2$ emissions, and NO$_X$ emissions as the three outputs.}
\label{Fig:ActivityPropelVehicleFromCrudeOil}
\end{figure}

The second diagram, called the Activity Diagram (ACT), is used to model a system's function in terms of its constituent activities. An example ACT is shown in Fig.~\ref{Fig:ActivityPropelVehicleFromCrudeOil}. In the context of this paper, an ACT contains (at a minimum):
\begin{itemize}
\item \textbf{Actions (or Activities)}: Represent specific tasks, operations, or processes within the system.  Note that the operations found in a BDD are often represented as actions in an ACT.  Similarly, there is no distinction between a process in a process flow diagram and an action in ACT.  
\item
 \textbf{Transitions}: Arrows indicate the flow from one action to another, showing the sequential or parallel execution of processes.
\item \textbf{Swim Lanes}: Horizontal or vertical partitions that separate the activities performed by different system components or roles, clarifying responsibilities.  The presence of an action within a swim lane also describes a capability.  
\end{itemize}
Together, the activity diagram depicts how the system's function is executed to reveal the system's behavior.  In the context of Fig. \ref{Fig:ActivityPropelVehicleFromCrudeOil}, an activity diagram entitled ``Propel Vehicle from Crude Oil" is used to describe the flow of processes that convert crude oil into distance travelled (km), NO$_x$ emissions, and CO$_2$ emissions.  These processes include: 
\begin{enumerate}
\item produce refined oil from crude oil,
\item generate electricity from refined oil,
\item transform electricity into distance travelled,
\item produce gasoline from crude oil, and
\item combust gasoline to the distance travelled.
\end{enumerate}
Additionally, vertical swim lanes are used to illustrate how each of these processes is allocated to the four types of resources shown in the BDD.  

\vspace{-0.2in}
\subsection{Hetero-functional Graph Theory for Life Cycle Analysis}\label{Sec:BackgroundHFGT}
\begin{figure}[htbp]
\centering
\includegraphics[width=\textwidth]{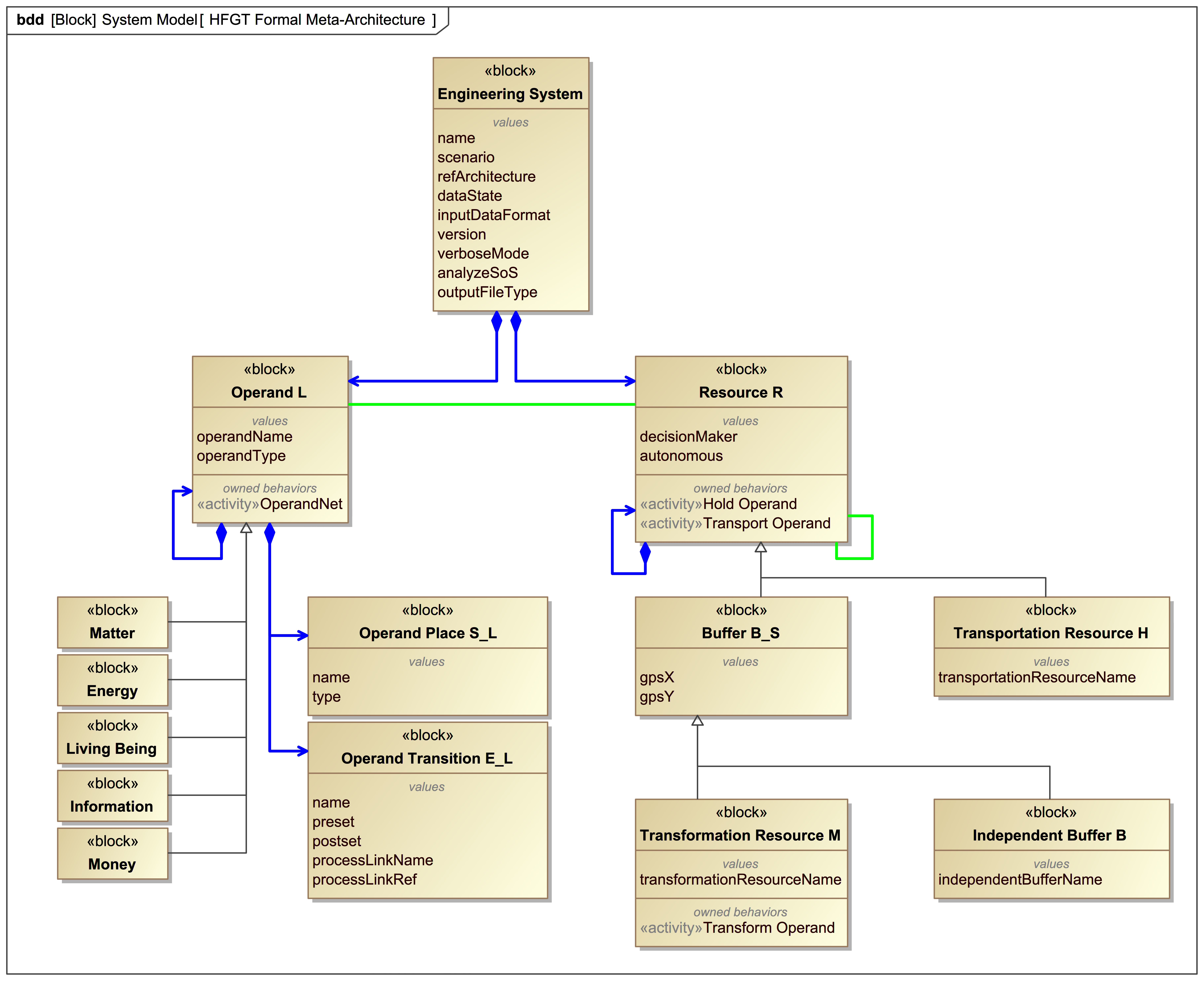}
\caption{A SysML Block Definition Diagram of the System Form of the Engineering System Meta-Architecture\cite{Schoonenberg:2019:ISC-BK04}.}
\label{Fig:LFESMetaArchitecture}
\end{figure}

While MBSE, and more specifically SysML, can graphically model large, complex systems, it does not have in-built functionality for conducting quantitative analysis.   Fortunately, hetero-functional graph theory (HFGT) offers an analytical approach for translating graphical SysML models into mathematical models that may be used for life cycle analysis.  As shown in Fig. \ref{Fig:LFESMetaArchitecture}, this translation requires the HFGT meta-architecture stated in SysML.  

The HFGT meta-architecture introduces several meta-elements whose definitions are formally introduced here:  
\begin{defn}[System Operand \cite{SE-Handbook-Working-Group:2015:00}]\label{Defn:SystemOperand}
An asset or object $l_i \in L$ that is operated on or consumed during the execution of a process.
\end{defn}
\begin{defn}[System Process\cite{Hoyle:1998:00,SE-Handbook-Working-Group:2015:00}]\label{Defn:D2SystemProcess}
An activity $p \in P$ that transforms or transports a predefined set of input operands into a predefined set of outputs. 
\end{defn}
\begin{defn}[System Resource \cite{SE-Handbook-Working-Group:2015:00}]\label{Defn:D3SystemResource}
An asset or object $r_v \in R$ that facilitates the execution of a process.  
\end{defn}
\noindent Importantly, these meta-elements are organized around the universal structure of human language.  Namely, system resources $R$ serve as subjects, system processes $P$ serve as predicates, and operands $L$ serve as objects within the predicates.  

The system resources $R=M \cup B \cup H$ are classified into transformation resources $M$, independent buffers $B$, and transportation resources $H$.  Additionally, the set of ``buffers" $B_S=M \cup B$ is introduced to support the discussion.  
\begin{defn}[Buffer\cite{Schoonenberg:2019:ISC-BK04,Farid:2022:ISC-J51}]\label{Defn:D4Buffer}
A resource $r_v \in R$ is a buffer $b_s \in B_S$ if it is capable of storing or transforming one or more operands at a unique location in space.  
\end{defn}
Equally important, the system processes $P = P_\mu \cup P_{\bar{\eta}}
$ are classified into transformation processes $P_\mu$ and refined transportation processes $P_\eta
$.  The latter arises from the simultaneous execution of one transportation process and one holding process.  Finally, hetero-functional graph theory emphasizes that resources are capable of one or more system processes to produce a set of ``capabilities"\cite{Schoonenberg:2019:ISC-BK04}.
\begin{defn}[Capability\cite{Schoonenberg:2019:ISC-BK04,Farid:2022:ISC-J51,Farid:2016:ISC-BC06}]\label{Defn:D5Capability}
An action $e_{wv} \in {\cal E}_S$ (in the SysML sense) defined by a system process $p_w \in P$ being executed by a resource $r_v \in R$.  It constitutes a subject + verb + operand sentence of the form: ``Resource $r_v$ does process $p_w$".  
\end{defn}
\noindent The highly generic and abstract nature of these definitions has allowed HFGT to be applied to numerous application domains, including electric power, potable water, wastewater, natural gas, oil, coal, multi-modal transportation, mass-customized production, and personalized healthcare delivery systems.  For a more in-depth description of HFGT, readers are directed to past works\cite{Schoonenberg:2019:ISC-BK04,Farid:2022:ISC-J51,Farid:2016:ISC-BC06}.

Returning to Fig. \ref{Fig:LFESMetaArchitecture}, the engineering system meta-architecture stated in SysML must be instantiated and ultimately transformed into the associated Petri net model. To that end, the positive and negative hetero-functional incidence tensors (HFIT) are introduced to describe the flow of operands through buffers and capabilities.  
\begin{defn}[The Negative 3$^{rd}$ Order Hetero-functional Incidence Tensor (HFIT) $\widetilde{\cal M}_\rho^-$\cite{Farid:2022:ISC-J51}]\label{Defn:D6 HFIT -ve}
The negative hetero-functional incidence tensor $\widetilde{\cal M_\rho}^- \in \{0,1\}^{|L|\times |B_S| \times |{\cal E}_S|}$  is a third-order tensor whose element $\widetilde{\cal M}_\rho^{-}(i,y,\psi)=1$ when the system capability ${\epsilon}_\psi \in {\cal E}_S$ pulls operand $l_i \in L$ from buffer $b_{s_y} \in B_S$.
\end{defn} 
\begin{defn}[The Positive  3$^{rd}$ Order Hetero-functional Incidence Tensor (HFIT)$\widetilde{\cal M}_\rho^+$\cite{Farid:2022:ISC-J51}]\label{Defn:D7 HFIT +ve}
 The positive hetero-functional incidence tensor $\widetilde{\cal M}_\rho^+ \in \{0,1\}^{|L|\times |B_S| \times |{\cal E}_S|}$  is a third-order tensor whose element $\widetilde{\cal M}_\rho^{+}(i,y,\psi)=1$ when the system capability ${\epsilon}_\psi \in {\cal E}_S$ injects operand $l_i \in L$ into buffer $b_{s_y} \in B_S$.
\end{defn}
\noindent These incidence tensors are straightforwardly ``matricized" to form the 2$^{nd}$ Order Hetero-functional Incidence Matrix $M = M^+ - M^-$ with dimensions $|L||B_S|\times |{\cal E}|$. Consequently, the supply, demand, transportation, storage, transformation, assembly, and disassembly of multiple operands in distinct locations over time can be described by an Engineering System Net and its associated State Transition Function\cite{Schoonenberg:2022:ISC-J50}.
\begin{defn}[Engineering System Net\cite{Schoonenberg:2022:ISC-J50}]\label{Defn:8 ESN}
An elementary Petri net ${\cal N} = \{S, {\cal E}_S, \textbf{M}, W, Q\}$, where
\begin{itemize}
\item $S$ is the set of places with size: $|L||B_S|$,
\item ${\cal E}_S$ is the set of transitions with size: $|{\cal E}|$,
\item $\textbf{M}$ is the set of arcs, with the associated incidence matrices: $M = M^+ - M^-$,
\item $W$ is the set of weights on the arcs, as captured in the incidence matrices,
\item $Q=[Q_B; Q_E]$ is the marking vector for both the set of places and the set of transitions. 
\end{itemize}
\end{defn}
\begin{defn}[Engineering System Net State Transition Function\cite{Schoonenberg:2022:ISC-J50}]\label{Defn:9 ESN-STF}
The  state transition function of the engineering system net $\Phi()$ is:
\begin{equation}\label{CH6:eq:PhiCPN}
Q[k+1]=\Phi(Q[k],U^-[k], U^+[k]) \quad \forall k \in \{1, \dots, K\}
\end{equation}
where $k$ is the discrete time index, $K$ is the simulation horizon, $Q=[Q_{B}; Q_{\cal E}
]$, $Q_B$ has size $|L||B_S| \times 1$, $Q_{\cal E}$ has size $|{\cal E}_S|\times 1$, the input firing vector $U^-[k]$ has size $|{\cal E}_S|\times 1$, and the output firing vector $U^+[k]$ has size $|{\cal E}_S|\times 1$.  
\begin{align}\label{Eq:ESNSTF1}
Q_{B}[k+1]&=Q_{B}[k]+{M}^+U^+[k]\Delta T-{M}^-U^-[k]\Delta T \\ \label{Eq:ESNSTF2}
Q_{\cal E}[k+1]&=Q_{\cal E}[k]-U^+[k]\Delta T +U^-[k]\Delta T
\end{align}
where $\Delta T$ is the duration of the simulation time step.  
\end{defn}
When the capabilities are assumed to occur instantaneously, $U^+[k]=U^-[k] \forall k$.  Then, Eq. \ref{CH6:eq:PhiCPN} collapses to triviality and Eq. \ref{Eq:ESNSTF1} becomes
\begin{align}\label{Eq:SimpleSTF}
Q_{B}[k+1]&=Q_{B}[k]+{M}U[k]
\Delta T 
\end{align}
Eq. \ref{Eq:SimpleSTF} then provides a means of conducting a life cycle analysis. More specifically, the marking vector $Q_B$ is partitioned into two groups:  the products $Y$ and the environmental aspects $E$. Niraj.G et al. \cite{Niraj.G:2025:00}.
\begin{align}
Q_B[
k] = \begin{bmatrix}
Y[k]\,\,;\,\,E[k]
\end{bmatrix}
\end{align}
Consequently, the hetero-functional incidence matrix can be similarly partitioned
\begin{align}
M = \begin{bmatrix}
A\,\,;\,\,B
\end{bmatrix}
\end{align}
Therefore, the engineering system net state transition function then becomes:
\begin{align}\label{Eq:SimpleSTFF}
\begin{bmatrix}
Y \\
E \end{bmatrix}[k+1] &= \begin{bmatrix}
Y \\
E \end{bmatrix}[k] + \begin{bmatrix}
A \\
B \end{bmatrix}U[k]
\Delta T 
\end{align}
Note that the number of process outputs in $Y$ is assumed to equal the number of processes in the firing vector $U$ and therefore the $A$ matrix is square and invertible.  In steady-state (rather than temporal) life cycle analysis, a simulation horizon of $K=2$ and a simulation time step of $\Delta T=1
$ are assumed.  Then the change in environmental aspects $\Delta E = E[k=2] - E[k=1]$ is solved in terms of the change of products $\Delta Y = Y[k=2] - Y[k=1]$.
\begin{align}
\Delta E = BA^{-1}\Delta Y
\label{Eq:SimpleLCA}
\end{align}

\section{Methodological Demonstration:  Spatio-Temporal Life Cycle Analysis of Electrolytic H2 Production in Australia}\label{Sec:LCA-AUS}

Given the background provided in the previous section, the spatio-temporal life cycle analysis is demonstrated for electrolytic hydrogen production in Australia.

\subsection{Model-Based Systems Engineering for Life Cycle Analysis: Block Definition Diagram}
\begin{figure}[H]
\centering
\includegraphics[width=1\linewidth]{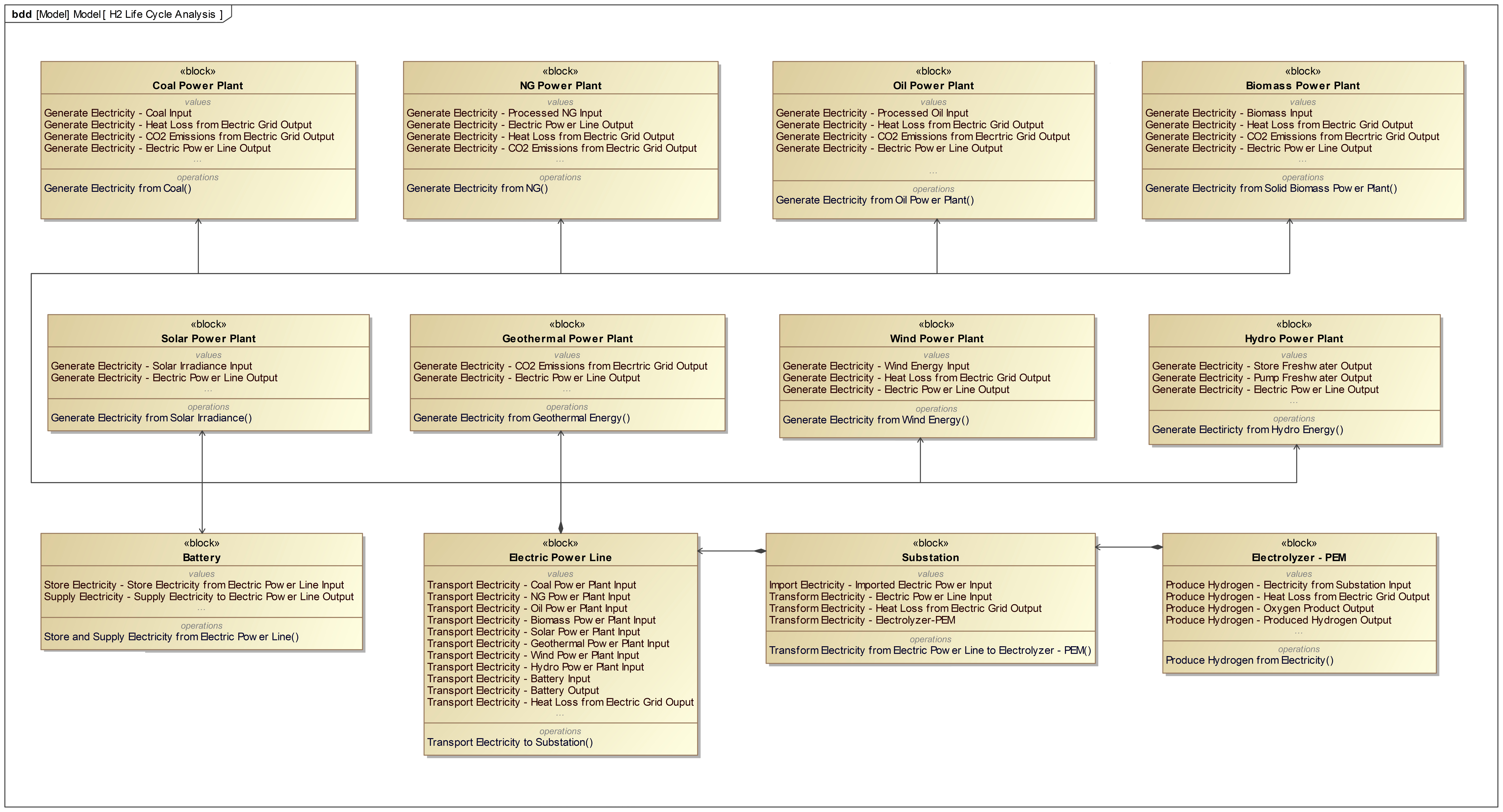}
\caption{A block definition diagram illustrating the Hydrogen Life Cycle System, presenting key processes involved in hydrogen production via grid electricity generated from various fuel types, each associated with distinct emissions profiles.}
\label{Fig:BDD_H2_LCA} 
\end{figure}

The Block Definition Diagram (BDD) in Fig. \ref{Fig:BDD_H2_LCA} entitled \emph{H$_2$ Life Cycle Analysis BDD} illustrates the system architecture of hydrogen production via various electricity sources in Australia. It models multiple types of generation facilities as blocks with their associated processes and emission factors: \emph{Coal Power Plant}, \emph{NG Power Plant}, \emph{Oil Power Plant}, \emph
{Biomass Power Plant}, \emph{Solar Power Plant}, \emph{Geothermal Power Plant}, \emph{Wind Power Plant}, and \emph{Hydro Power Plant}. Additionally, these generation facilities feed into a \emph{Battery}, \emph{Electric Power Line}, and \emph{Substation}, which manage electricity storage, transportation, and transformation, respectively. Finally, the \emph{PEM Electrolyzer} uses electricity to produce hydrogen through the electrolysis of water.  Each block clearly outlines its inputs (e.g., fuel, water), outputs (e.g., electricity, emissions, hydrogen), and the associated transformation operations, reflecting a comprehensive systems engineering approach to modeling the hydrogen life cycle from generation to end use.

\subsection{Model-Based Systems Engineering for Life Cycle Analysis: Activity Diagram}
\begin{figure}[H]
\centering
\includegraphics[width=1\linewidth]{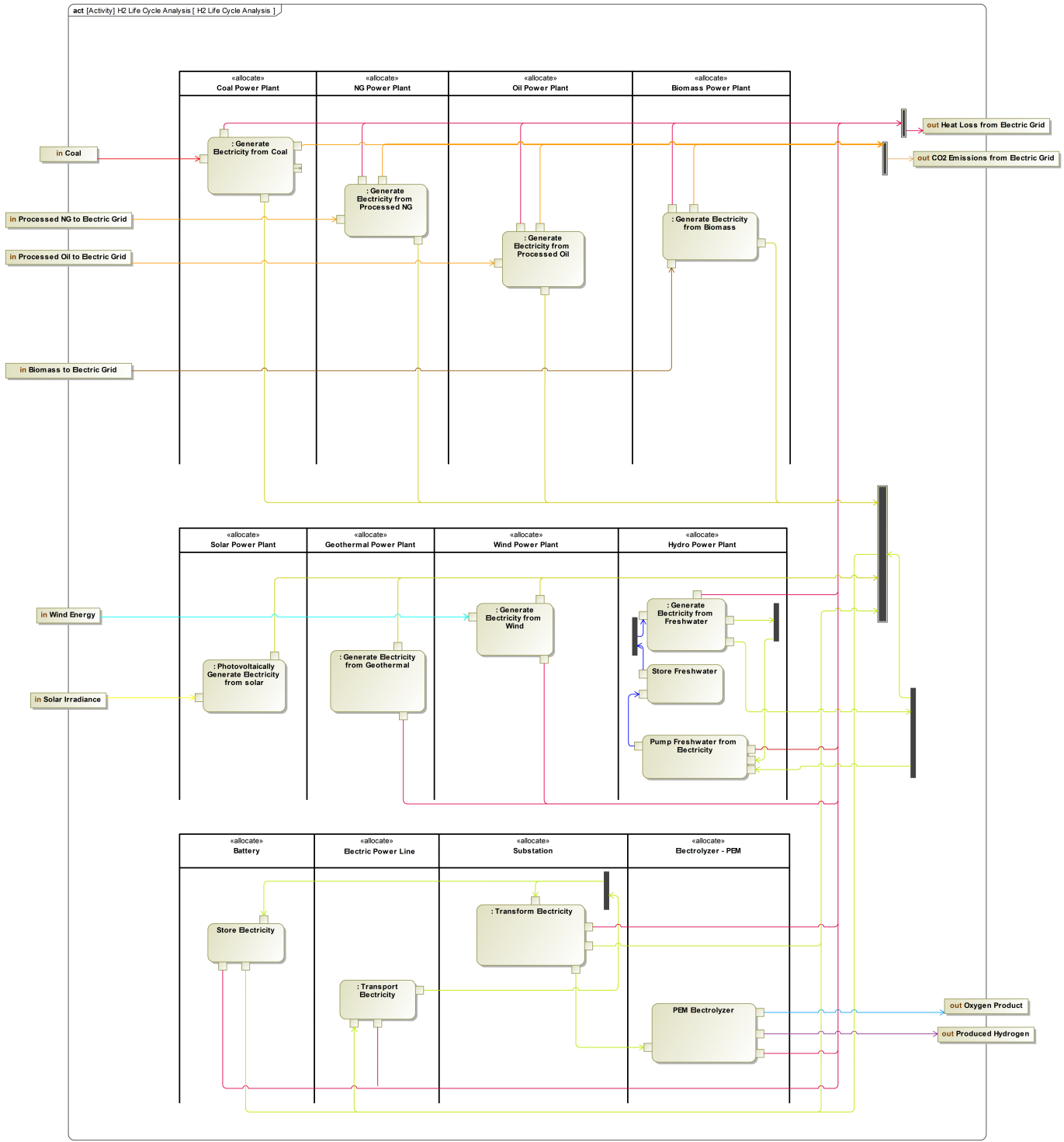}
\caption{Activity diagram illustrating the hydrogen life cycle system, capturing the flow of energy and materials from diverse electricity generation facilities to hydrogen production.}
\label{Fig:ACT_H2_LCA}
\end{figure}

The Activity Diagram in Fig. \ref{Fig:ACT_H2_LCA} entitled \emph{H$_2$ Life Cycle Analysis ACT} is the SysML equivalent of a process flow diagram in process-based life cycle analysis.  It illustrates the flow of energy conversion processes involved in producing hydrogen from electricity generated from various power sources in Australia.  Various energy sources — including coal, natural gas, oil, biomass, solar irradiance, wind energy, geothermal energy, and hydroelectric power are converted into electricity through their respective generation facilities.  These processes are represented using activities such as “Generate Electricity” and “Photovoltaically Generate Electricity from Solar”.  These activities are connected by flows of matter and energy inputs (e.g., water, biomass, wind, irradiance) and outputs (e.g., electricity, CO$_2$ emissions, heat loss). Each of these activities is also assigned to its respective resource (i.e. facility) via a vertical swim lane.  The diagram culminates in the production of hydrogen and oxygen as outputs, representing the end-use utility of the system. Overall, this activity diagram offers a comprehensive view of the operational interactions and transformations necessary for sustainable hydrogen production.

\subsection{Hetero-functional Graph Theory for Life Cycle Analysis} \label{Sec.III-C HFGT}
The elaboration of the SysML block definition diagram and activity diagram allows for the straightforward application of hetero-functional graph theory.  More specifically, each definition in Sec. \ref{Sec:Background} is followed.

\begin{trivlist}
\item \textbf{System Operands:} Following Defn.~\ref{Defn:SystemOperand}, the system operands are identified from the inputs and outputs shown in Fig.~\ref{Fig:ACT_H2_LCA}. Consequently,

\begin{align}\nonumber
L = \{ &\mbox{Coal, Processed NG, Processed Oil, Biomass, Solar Irradiance, Geothermal Power, Wind Energy, } \\\nonumber 
&\mbox{Hydro Power, Heat Loss from Electric Grid, CO\textsubscript{2} Emissions from Electric Grid, } \\\nonumber 
&\mbox{ Oxygen Product, Produced Hydrogen} \} 
\end{align}\vspace{-0.1in}

\item \textbf{System Processes:} Following Defn.~\ref{Defn:D2SystemProcess}, the system processes are identified from the activities shown in Fig.~\ref{Fig:ACT_H2_LCA}. Consequently,
\begin{align}\nonumber
P = \{ &\mbox{Generate Electricity from Coal, Generate Electricity from Processed NG, } \nonumber \\
&\mbox{Generate Electricity from Processed Oil, Generate Electricity from Biomass,  } \nonumber \\
&\mbox{Generate Electricity from Solar Irradiance, Generate Electricity from Geothermal, } \nonumber \\
&\mbox
{Generate Electricity from Wind, Generate Electricity from Freshwater, } \nonumber \\
&\mbox{Pump Freshwater from Electricity, Transform Electricity, Import Electricity, }\nonumber \\
&\mbox{Produce Hydrogen from Electricity, Store Electricity, Transport Electricity, Store Freshwater} \}
\end{align}
\noindent Note that the last three processes are transportation processes, while the rest are transformation processes.  
\item \textbf{System Resources:} Following Defn. ~\ref{Defn:D3SystemResource}, the system resources are identified equivalently from the blocks in Fig. ~\ref{Fig:BDD_H2_LCA} or the swim lanes in Fig.~\ref{Fig:ACT_H2_LCA}.  Consequently, 
\begin{align}\nonumber
R = \{ &\mbox{Coal Power Plant, NG Power Plant, Oil Power Plant, Biomass Power Plant, Solar Power Plant, } \nonumber\\
&\mbox {Geothermal Power Plant, Wind Power Plant, Hydro Power Plant, Battery, Substation} \nonumber \\
&\mbox {Electrolyzer - PEM, Electric Power Line}\}
\end{align}
\noindent Note that for clarity of exposition, and without loss of generality, each type of resource shown in Fig.~\ref{Fig:BDD_H2_LCA} and Fig.~\ref{Fig:ACT_H2_LCA} is assumed to be instantiated exactly once. Therefore, if a given region has more than one resource of a specific type, this analysis aggregates them into a single resource of the same type.  
\item \textbf{Buffer:} Following Defn.~\ref{Defn:D4Buffer}, the buffers are the system resources that have a fixed point location in space.  This includes all of the above resources except the electric power line.  
\item \textbf{Capability:} Following Defn.~\ref{Defn:D5Capability}, the system capabilities are identified equivalently from the operations associated with blocks in Fig.~\ref{Fig:BDD_H2_LCA} or the activities associated with swimlanes in Fig.~\ref{Fig:ACT_H2_LCA}.  Consequently, 
\begin{align}\nonumber
{\cal E}_S = \{ 
&\mbox{Coal Power Plant generates Electricity from Coal,} \nonumber\\
&\mbox{NG Power Plant generates Electricity from Processed NG,} \nonumber\\
&\mbox{Oil Power Plant generates Electricity from Processed Oil,} \nonumber\\
&\mbox{Biomass Power Plant generates Electricity from Biomass,} \nonumber\\
&\mbox{Solar Power Plant generates Electricity from Solar Irradiance,} \nonumber\\
&\mbox{Geothermal Power Plant generates Electricity from Geothermal Power,} \nonumber\\
&\mbox{Wind Power Plant generates Electricity from Wind Energy,} \nonumber\\
&\mbox{Hydro Power Plant generates Electricity from Hydro Power,} \nonumber\\
&\mbox{Pumping Station generates Withdrawn Freshwater using Electricity,} \nonumber\\
&\mbox{Reservoir stores Withdrawn Freshwater,} \nonumber\\
&\mbox{Battery stores Electricity,} \nonumber\\
&\mbox{Battery discharges Stored Electricity to Grid,} \nonumber\\
&\mbox{Substation transforms Electricity,} \nonumber\\
&\mbox{Grid imports Electricity,} \nonumber\\
&\mbox{Electric Power Line transports Electricity,} \nonumber\\
&\mbox{PEM Electrolyzer generates Hydrogen and Oxygen from Electricity and Water} 
\}
\end{align}

\item \textbf{Hetero-functional Incidence Matrix:}
\noindent The hetero-functional incidence tensor $\widetilde{\cal M}_
\rho$ is determined from Defn. \ref{Defn:D6 HFIT -ve} and \ref{Defn:D7 HFIT +ve} and then matricized.  The resulting matrix has a size of $|L||B_S|x|{\cal E}_S|$ and its values must be inferred from  Fig. \ref{Fig:BDD_H2_LCA} and Fig. \ref{Fig:ACT_H2_LCA}.  Note that the hetero-functional incidence matrix has rows filled with zeros because not all combinations of operand and buffer are necessary.  When these zero-filled rows are eliminated, the resulting hetero-functional incidence matrix $M$ becomes:

\[
M = \begin{bmatrix}
A \\ B
\end{bmatrix} = 
\left[
\begin{array}{rrrrrrrrrrrrrrr}
\mathcal{E}_{S_1} & \mathcal{E}_{S_2} & \mathcal{E}_{S_3} & \mathcal{E}_{S_4} & \mathcal{E}_{S_5} & \mathcal{E}_{S_6} & \mathcal{E}_{S_7} & \mathcal{E}_{S_8} & \mathcal{E}_{S_9} & \mathcal{E}_{S_{10}} & \mathcal{E}_{S_{11}} & \mathcal{E}_{S_{12}} & \mathcal{E}_{S_{13}} \\\hline
 -4.6 &     0 &     0 &   0 &   0 &     0 &     0 &   0 &   0 &   0 &   0 &   0 &  0 \\
    0 &  -1.3 &     0 &   0 &   0 &     0 &     0 &   0 &   0 &   0 &   0 &   0 &  0 \\
    0 &     0 &  -2.7 &   0 &   0 &     0 &     0 &   0 &   0 &   0 &   0 &   0 &  0 \\
    0 &     0 &     0 &   0 &   0 &  -3.7 &     0 &   0 &   0 &   0 &   0 &   0 &  0 \\
    0 &     0 &     0 &   0 &   0 &     0 &     0 &   0 &   0 &   0 &   0 &   0 &  0 \\
 30.3 &     0 &     0 &   0 &   0 &     0 &     0 &   0 &   0 &   0 &   0 &   0 &  0 \\
    0 &  16.7 &     0 &   0 &   0 &     0 &     0 &   0 &   0 &   0 &   0 &   0 &  0 \\
    0 &     0 &  32.3 &   0 &   0 &     0 &     0 &   0 &   0 &   0 &   0 &   0 &  0 \\
    0 &     0 &     0 &   1 &   0 &     0 &     0 &   0 &   0 &   0 &   0 &   0 &  0 \\
    0 &     0 &     0 &   0 &   1 &     0 &     0 &   0 &   0 &   0 &   0 &   0 &  0 \\
    0 &     0 &     0 &   0 &   0 &  28.6 &     0 &   0 &   0 &   0 &   0 &   0 &  0 \\
    0 &     0 &     0 &   0 &   0 &     0 &     1 &   0 &   0 &   0 &   0 &   0 &  0 \\
    0 &     0 &     0 &   0 &   0 &     0 &     0 &   1 &   0 &   0 &   0 &   0 &  0 \\
    0 &     0 &     0 &   0 &   0 &     0 &     0 &   0 &  90 &   0 &   0 & -90 &  0 \\
    0 &     0 &     0 &   0 &   0 &     0 &     0 &   0 &   0 & -90 &  90 &   0 &  0 \\
    0 &     0 &     0 &   0 &   0 &     0 &     0 &   0 &   0 &   0 &   0 &   0 & -52.5 \\
    0 &     0 &     0 &   0 &   0 &     0 &     0 &   0 &   0 &   0 &   0 &   0 &  1 \\\hline
-20.3 &  -6.7 & -22.3 &   0 &   0 & -18.6 &     0 &   0 &   0 &   1 &   1 &   0 &  1 \\
   820&   490 &   650 &   0 &   0 &   230 &     0 &   0 &   0 &   0 &   0 &   0 &  0 \\
    0 &     0 &     0 &   0 &   0 &     0 &     0 &   0 &   0 &   0 &   0 &   0 &  8 \\
\label{incidence_matrix}
\end{array}
\right]
\]
\begin{figure}[H]
\centering
\includegraphics[width=1\linewidth]{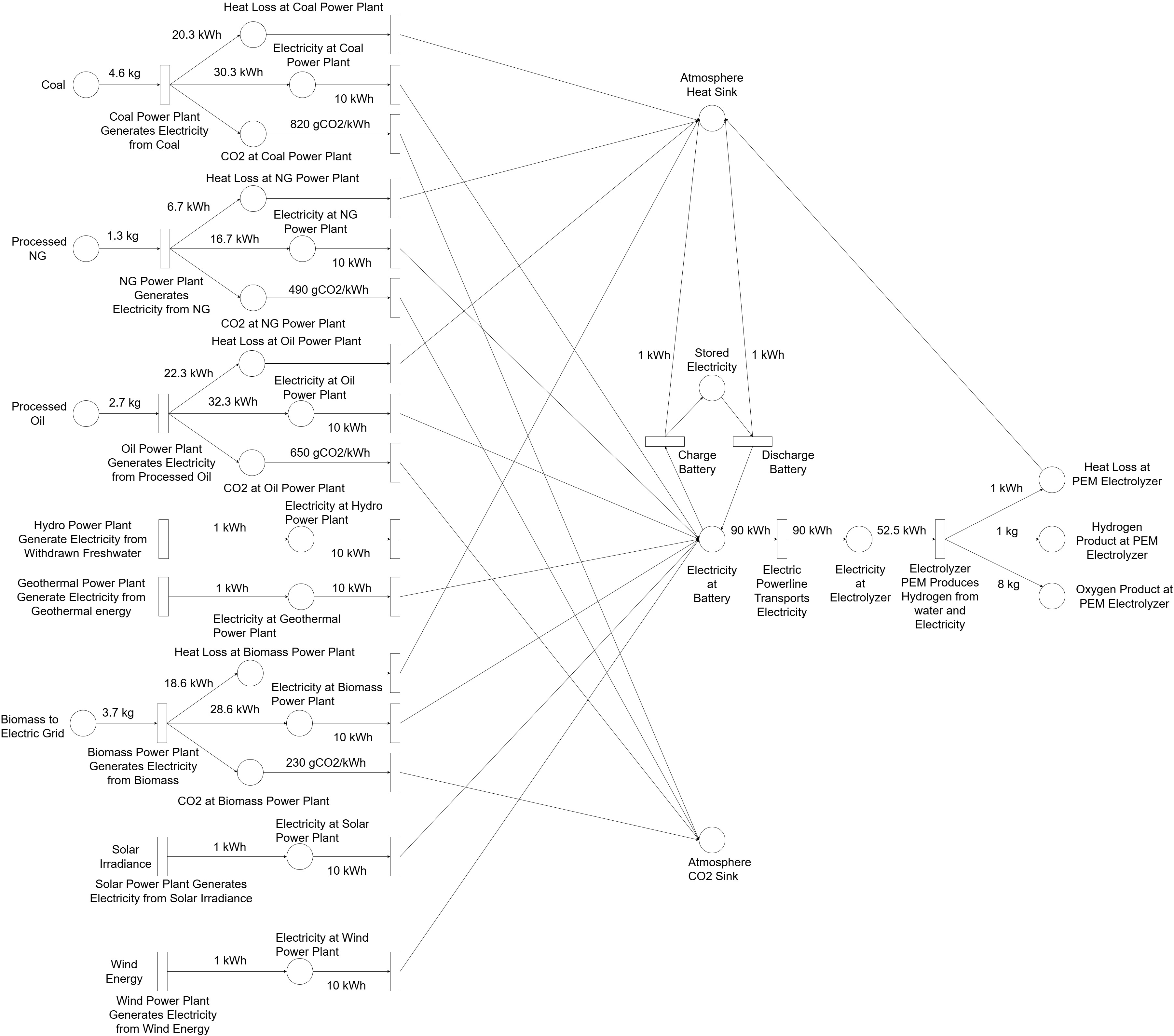}
\caption{Engineering System Net (Petri Net) representation of the hydrogen production life cycle via grid electricity in Australia. The model captures the complete flow of energy, materials, and emissions from primary fuel inputs (e.g., coal, natural gas, oil, biomass, and renewables) through electricity generation, transmission, and battery storage to hydrogen production via a PEM electrolyzer. Each transition denotes a process, while places represent intermediate energy or material states, including heat losses, emissions to the atmosphere, and hydrogen/oxygen outputs. Energy and carbon flows are quantified for each process, enabling temporally resolved life cycle assessment using the Hetero-functional Graph Theory (HFGT) framework.}
\label{fig:Petri_Net}
\end{figure}

\item The values in the Petri Net incidence matrix were derived using hourly electricity generation data from Electricity Maps, which was analyzed in the attached Excel sheet. Each transition weight in the matrix corresponds to the energy contribution (in kWh) of specific fuel types or conversion processes across 24 hours, while emission factors were assigned based on IPCC and GREET model standards. The bottom row aggregates the carbon emission intensity (gCO$_2$eq/kWh) per fuel type to compute the lifecycle carbon flow through the Petri Net system. 

\item \textbf{Engineering System Net:}  Finally, following Defn.~\ref{Defn:8 ESN}, the engineering system net is constructed and depicted in Fig.~\ref{fig:Petri_Net}.  Naturally, its state transition function follows Defn. \ref{Defn:9 ESN-STF}.  The places of the engineering system net is partitioned into process outputs $Y$ and environmental aspects $E$.  
\begin{align}
\Delta Q_{B}&={M}U \\\label{Eq:PBLCA}
\begin{bmatrix}
Y \\ E
\end{bmatrix} &= 
\begin{bmatrix}
A \\ B
\end{bmatrix}X
\end{align}
\end{trivlist}

\subsection{Validation of HFGT Life Cycle Analysis}

To evaluate the accuracy of the Hetero-functional Graph Theory (HFGT) based life cycle analysis methodology developed in this study, a comparative validation was conducted using publicly available data from Electricity Maps \cite{electricitymaps_datasets}. This platform reports hourly carbon intensity values derived from power generation mixes and standardized emission factors for different energy sources.  The carbon intensity data, emission factors, and generation breakdowns were downloaded for a representative 24-hour period. These inputs were used to reconstruct the hourly carbon intensity using the formulation described in Section \ref{Sec.III-C HFGT}. Specifically, each generation source was multiplied by its respective emission factor, and the weighted average carbon intensity was computed on an hourly basis.

\begin{figure}[H]
\centering
\includegraphics[width=\textwidth]{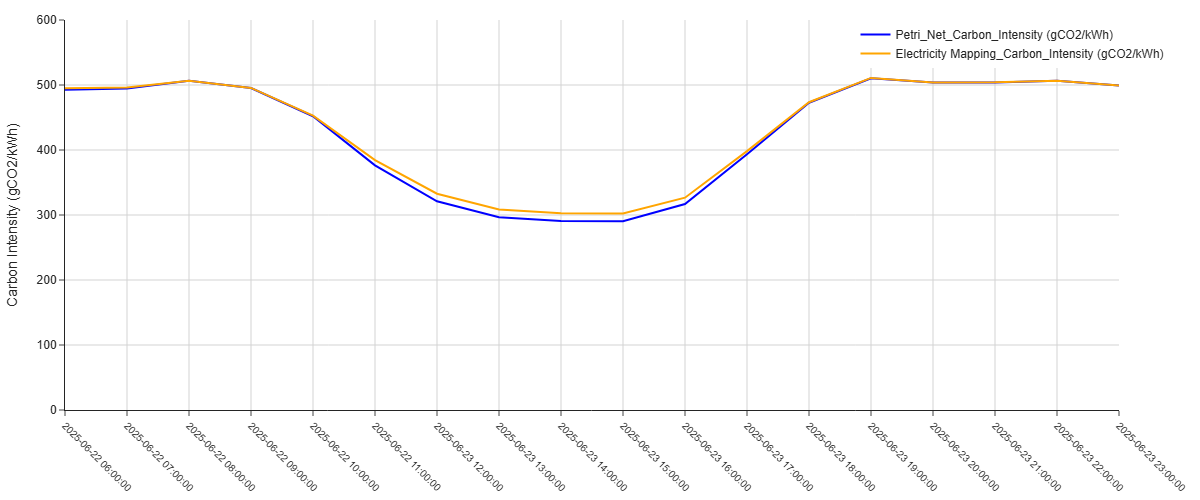}
\caption{Comparison of hourly carbon intensity values obtained using the proposed Petri Net-based HFGT life cycle analysis method and the reported values from Electricity Maps for a representative 18-hour period. The close alignment of the two curves demonstrates the validity and robustness of the proposed approach. Minor discrepancies are attributable to rounding artifacts and differences in data resolution.}
\label{fig:validation_comparison}
\end{figure}

The reconstructed results obtained through the HFGT-Petri Net framework closely align with those reported by Electricity Maps \cite{electricitymaps_datasets}. A comparison of both time series is shown in Figure~\ref{fig:validation_comparison}. It illustrates an 18-hour comparative analysis of hourly carbon intensity values derived from the Petri net in Fig. \ref{fig:Petri_Net} alongside those provided by Electricity Maps \cite{electricitymaps_datasets}. Minor deviations were observed, which are attributed to rounding artifacts during data acquisition.  The study demonstrates that the Engineering System Net, developed from the SysML architecture Fig. \ref{Fig:LFESMetaArchitecture} and incidence matrix \ref{incidence_matrix}, correctly calculates the temporal fluctuations of actual grid emissions. This coherence in results validates the model assumptions including operand flows, process weights, and transition logic. Consequently, the MBSE-HFGT based approach to dynamic life cycle assessment (LCA) provides a robust basis for its use in the spatio-temporal modeling of hydrogen production scenarios discussed in \ref{Sec:CaseStudy}.  

\newpage
\section{Full Demonstration:  Spatio-Temporal Life Cycle Analysis of Electrolytic H2 Production in Australia}
\label{Sec:CaseStudy} 
This section extends the preliminary demonstration of the previous section to a year-long life cycle analysis of electrolytic hydrogen production across five states in Australia:  New South Wales, Queensland, South Australia, Tasmania, and Victoria. The analysis draws upon generation data and carbon intensities of generation from Electricity Mapping\cite{electricitymaps_datasets}, to calculate life cycle carbon emissions.  It also pulls electricity pricing data from AEMO\cite{aemo_nem_data_dashboard}. The electrolyzer considered for this analysis is a 52.5 kWh/Kg H$_2$ PEM electrolyzer, which uses electricity to produce hydrogen through the electrolysis of water. The section is organized as follows.  Sec. \ref{Sec:AusCharacteristics} describes the spatio-temporal characteristics of life cycle carbon intensities and electricity prices in Australia.  The section then discusses and studies the outcome of three different hydrogen production scenarios.  Sec. \ref{Sec:BaselineScenario} studies a baseline scenario with hydrogen production at a fixed rate.  Sec. \ref{Sec:VariableProduction} studies a variable production scenario that produces hydrogen when the grid is relatively ``green".  Finally, Sec. \ref{Sec:SmartProduction} studies a ``smart production scenario" that exploits tax incentives.

\subsection{Spatio-Temporal Characteristics of Life Cycle Carbon Intensity and Electricity Prices in Australia}\label{Sec:AusCharacteristics}
\begin{figure}[H]
\centering
\includegraphics[width=0.9\linewidth]{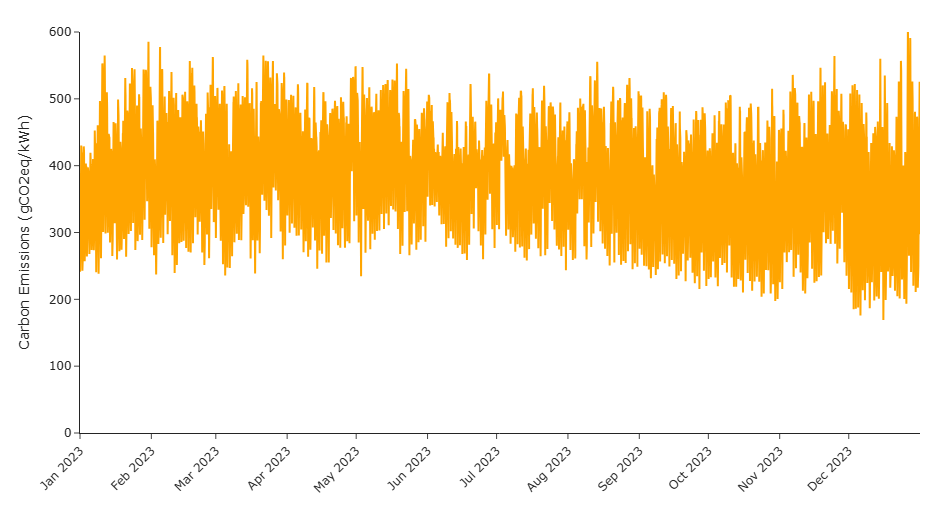}
\vspace{-0.2in}
\caption{Hourly time series of life cycle carbon intensity of electricity production in Australia in 2023. \cite{electricitymaps_datasets}}
\label{fig:2023_CO2_Emissions_Australia}
\end{figure}

To begin, Fig. \ref{fig:2023_CO2_Emissions_Australia} shows the hourly life cycle carbon intensity (LCA CO2eq) for electricity production across the Australian grid in the 2023 database. Electricity Mapping, treating all five states, New South Wales, Queensland, South Australia, Tasmania, and Victoria as a single interconnected grid. It illustrates the significant temporal variability in carbon intensity throughout the year, driven by various fluctuations such as regional energy mixes, demand patterns, and renewable generation availability. Also, reveals that an hourly temporal resolution is critical for identifying low-carbon windows for the electrolyzer operation. Thus, Fig. \ref{fig:2023_CO2_Emissions_Australia} serves as the foundational layer for the three hydrogen production scenarios analyzed in this study.  

\begin{figure}[H]
\centering
\includegraphics[width=0.9\linewidth]{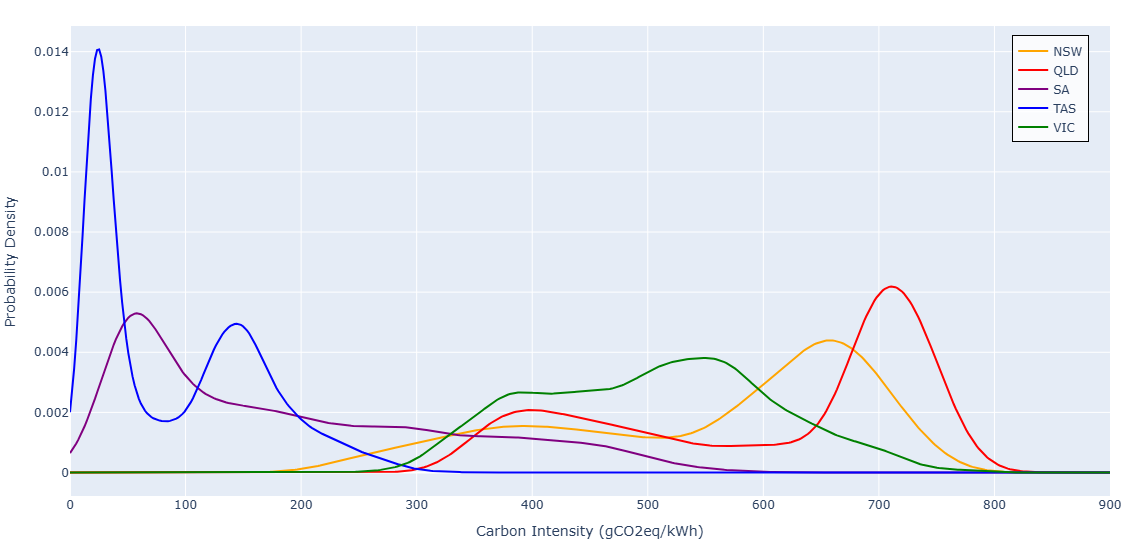}
\vspace{-0.1in}
\caption{Probability density function of life cycle carbon intensity (LCA CO$_2$eq) of electricity production for a.) New South Wales (orange), b.) Queensland (red), c.) South Australia (purple), d.) Tasmania (blue), e.) Victoria (green) in 2023. \cite{electricitymaps_datasets}}
\label{fig: PDFs_CO2eq}
\end{figure}

Additionally, Fig. \ref{fig: PDFs_CO2eq} reveals that the temporal variability of life cycle carbon intensities shown in Fig. \ref{fig:2023_CO2_Emissions_Australia} has spatial variability as well.  Fig. \ref{fig: PDFs_CO2eq} depicts the probability density function (PDFs) of life cycle carbon intensity (LCA CO2eq) of electricity production for each of the five Australian states in 2023. It highlights the distinct emissions associated with each state's grid. Tasmania exhibits a sharply peaked distribution at lower carbon intensities, reflecting its dominance of renewable sources such as hydroelectricity. In contrast, states like Queensland, Victoria, and New South Wales have broader distributions towards higher carbon intensities due to their greater dependency on coal and natural gas. These PDFs provide a statistical basis for understanding the likelihood of encountering low-carbon electricity at each hour in that state, which informs the timing of periods of hydrogen production with lower grid emissions. 

\begin{figure}[H]
\centering
\includegraphics[width=0.9\linewidth]{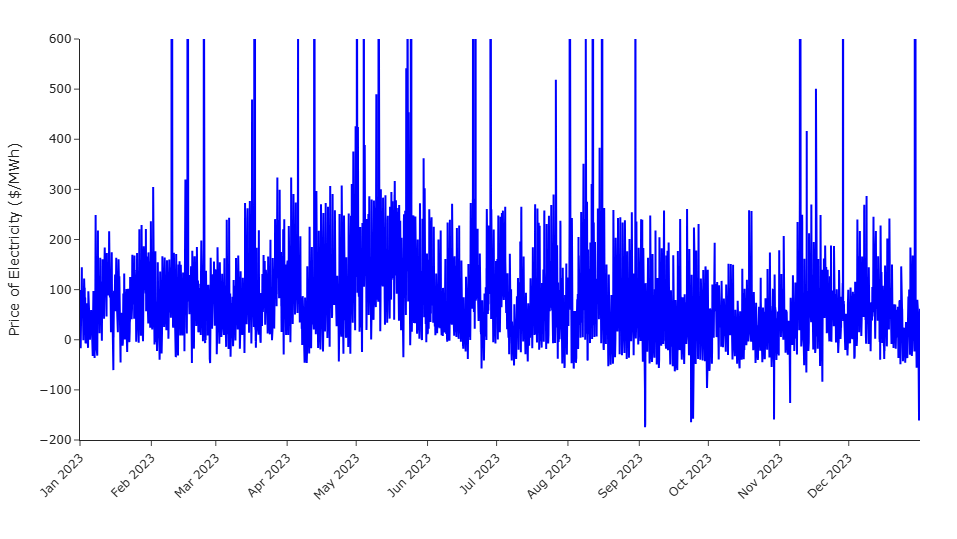}
\vspace{-0.2in}
\caption{Hourly time series of electricity prices in Australia in 2023\cite{aemo_nem_data_dashboard}.}
\label{fig:2023_Electricity_Price_Australia}
\end{figure}

Turning away from life cycle carbon intensities, Figure \ref{fig:2023_Electricity_Price_Australia} presents the hourly electricity price (\$AUD/MWh) time series for the entire Australian National Electricity Market for the full year of 2023 \cite{aemo_nem_data_dashboard}. It reveals frequent highs and lows in the price, including both positive and negative spot pricing events, indicating relative scarcity of supply or demand, respectively.  Furthermore, the variable nature of renewable energy contributions in the grid causes low prices when the renewable energy is at its peak, increasing the supply beyond the market demand.  This real-time price data is crucial for dynamic hydrogen production as it allows for cost minimization by shifting the electrolyzer operations to low-price hours. When considered with the data shown in Figs. \ref{fig:2023_CO2_Emissions_Australia} and \ref{fig: PDFs_CO2eq}, this data enables the formulation of strategies that will align the hydrogen production with both low-emissions and low-cost periods.  

\begin{figure}[H]
\centering
\includegraphics[width=0.9\linewidth]{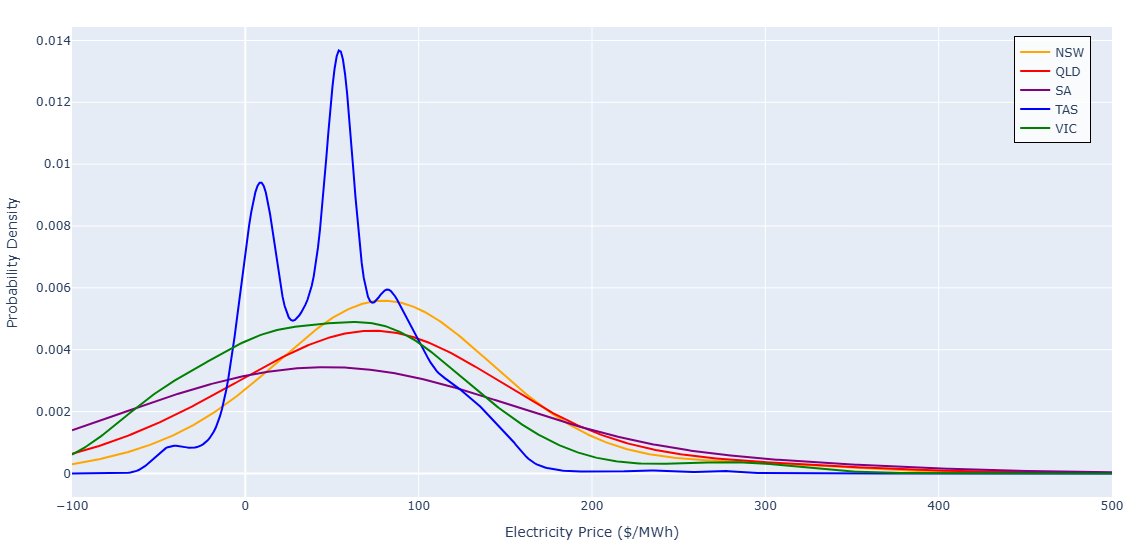}
\caption{Probability density function of electricity prices for a.) New South Wales (orange), b.) Queensland (red), c.) South Australia (purple), d.) Tasmania (blue), e.) Victoria (green) in 2023 \cite{aemo_nem_data_dashboard}.}
\label{fig:PDFs_electricity}
\end{figure}

Additionally, Fig. \ref{fig:PDFs_electricity} reveals that the temporal variability of electricity prices shown in Fig. \ref{fig:2023_Electricity_Price_Australia} has spatial variability as well.  Fig. \ref{fig:PDFs_electricity} presents the probability density functions (PDFs) of hourly electricity prices in 2023 for each of the five Australian states. It provides a statistical perspective on price behavior throughout the year, giving insights into the frequency and duration of low-cost electricity associated with each state's grid. South Australia and Tasmania display broader distributions with frequent low or even negative prices, reflecting the high renewable energy generation. Whereas New South Wales, Queensland, and Victoria exhibit more stable but generally higher average prices. When compared with the carbon intensity PDFs in Fig. \ref{fig: PDFs_CO2eq}, these electricity price distributions allow for the identification of optimal times where both environmental and economic performance can be maximized. Overall, these insights form the basis for the three scenarios analyzed in the following sections.   

\subsection{Baseline Scenario: Fixed Hydrogen Production}\label{Sec:BaselineScenario}

\begin{figure}[H]
\centering
\includegraphics[width=0.9\linewidth]{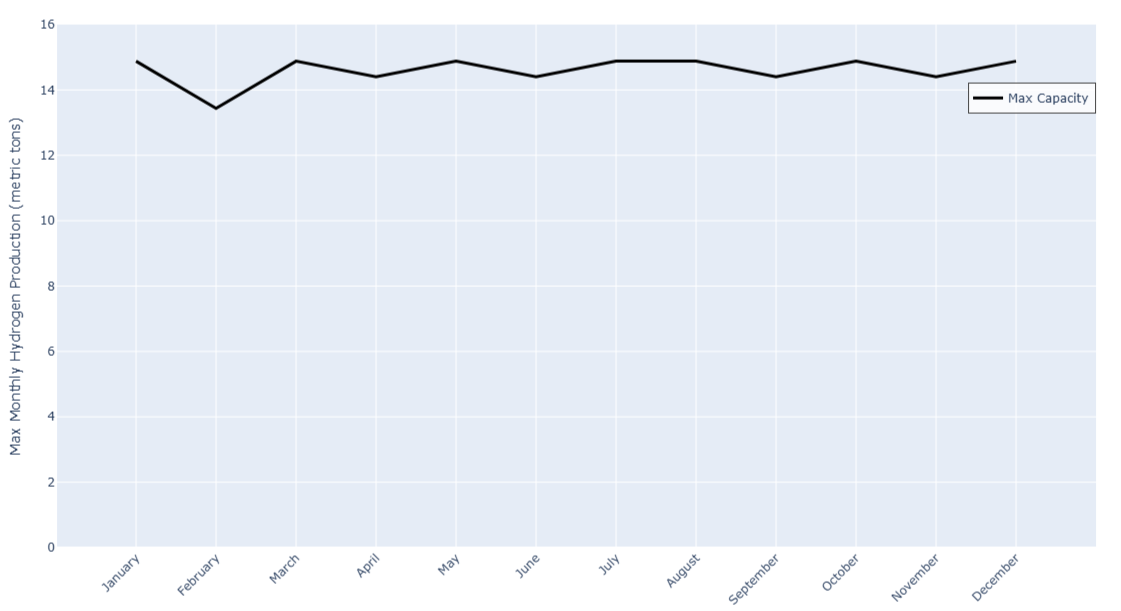}
\caption{Monthly Maximum Hydrogen Production Capacity (2023), assuming continuous operation of the electrolyzer at 20 kg H$_2$/hour (i.e., 24 hours/day for each month). Variations across months reflect differences in the number of days, leading to a maximum output ranging from 13.44 to 14.88 metric tons.}
\label{Fig:MaxProduction}
\end{figure}

In this baseline scenario, the electrolyzer operates on a traditional operational strategy for electrolytic hydrogen production at a constant and maximum rate of 20 kg/hour throughout the year 2023, regardless of variations in grid carbon intensity or electricity price in all five states of Australia. When this maximum production rate is summed, it results in the maximum monthly production values shown in Fig. \ref{Fig:MaxProduction}. Naturally, the maximum output ranges from 13.44 to 14.88 metric tons depending on the number of days in each month.  On this basis, the life cycle carbon emissions are calculated following the method described in Sec. \ref{Sec:Background} and demonstrated in Sec. \ref{Sec:LCA-AUS}.  Similarly, the cost of hydrogen production is derived by applying the fixed electricity consumption of 52.5 kWh per kg H$_2$ to the hourly electricity prices from the AEMO database, and an additional operational cost of \$1.96/the per kg of Hydrogen.  

\begin{figure}[H]
\centering
\includegraphics[width=0.9\linewidth]{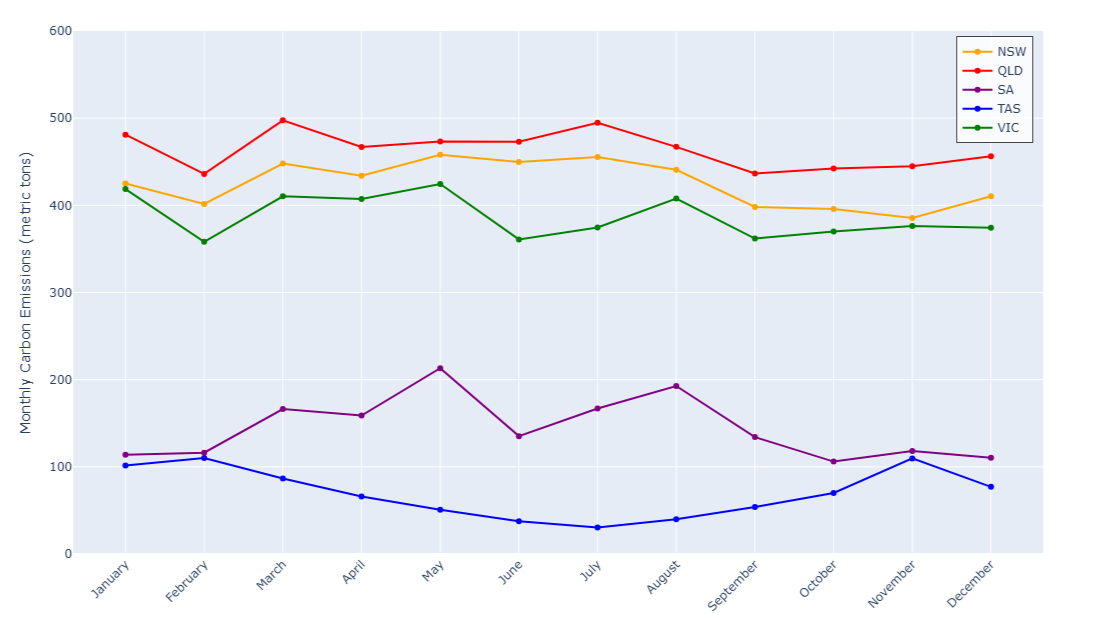}
\caption{Baseline Scenario - 20 kg H$_2$/hr for all hours in 2023 for a.) New South Wales (orange), b.) Queensland (red), c.) South Australia (purple), d.) Tasmania (blue), e.) Victoria (green) in 2023.  Monthly time series of life cycle carbon emissions in metric tons.}
\label{fig:CO2_output_all_months_S1}
\end{figure}

Figure \ref{fig:CO2_output_all_months_S1} illustrates the monthly life cycle carbon emissions in metric tons of CO$_2$ equivalent resulting from the baseline scenario of fixed hydrogen production across five states.  As seen, emissions vary spatially and seasonally.  The emissions for each state show the variation in the grid's generation mix.  As expected, Tasmania and South Australia show consistently low values due to its high penetration of renewable energy.  Interestingly, Tasmania's life cycle carbon emissions are least during the winter months when hydroelectric power is most plentiful.  In contrast, South Australia's life cycle carbon emission are least during the summer months when solar power is most plentiful.  Finally, Queensland, New South Wales, and Victoria record higher emissions due to a greater reliance on fossil fuels.  

\begin{figure}[H]
\centering
\includegraphics[width=0.9\linewidth]{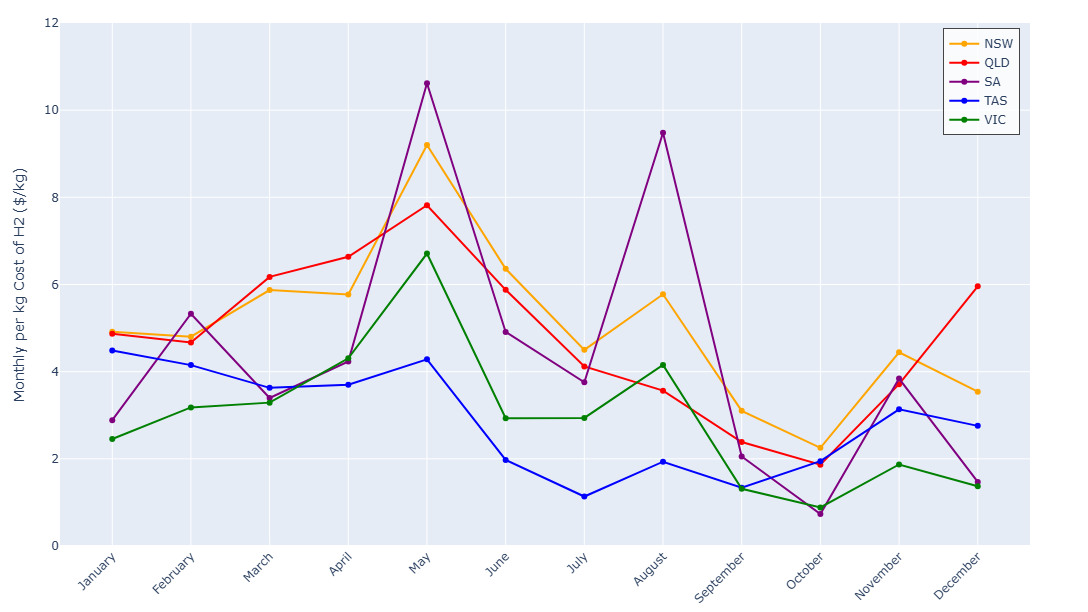}
\caption{Baseline Scenario - 20 kg H$_2$/hr for all hours in 2023 for a.) New South Wales (orange), b.) Queensland (red), c.) South Australia (purple), d.) Tasmania (blue), e.) Victoria (green) in 2023.  Monthly time series of per unit H$_2$ cost in $\$/kg$.}
\label{fig:Per_kg_h2_Cost_S1}
\end{figure}

Fig. \ref{fig:Per_kg_h2_Cost_S1} complements Fig. \ref{fig:CO2_output_all_months_S1} by showing the corresponding monthly per unit hydrogen production costs in (AUD/kg H$_2$).  Again, it reveals significant cost variations across both time and geography. Cost advantages are occasionally observed in regions like South Australia and Tasmania, which experience more frequent periods of low or negative electricity pricing. Together, these figures emphasize the limitations of fixed-rate production in capturing economic and environmental efficiencies.  Nevertheless, a fixed-rate production scenario establishes a benchmark for evaluating time-responsive production strategies (in the following scenarios).

\subsection{Variable Production Scenario 1: Produces H\texorpdfstring{$_2$}{2} When the Grid is Green}\label{Sec:VariableProduction}

\begin{figure}[H]
\centering
\includegraphics[
width=0.9\linewidth]{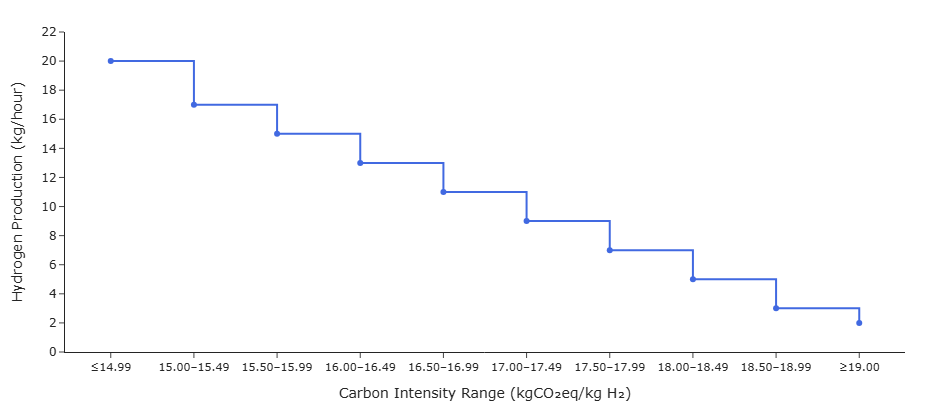}
\caption{Production Rule for Variable Production Scenario 1:  Depending on the life cycle carbon intensity of a given hour, the associated amount of H$_2$ is produced. }
\label{fig:Production_Rule_H2}
\end{figure}

This second scenario introduces a dynamic operational strategy where hydrogen is produced selectively during periods of low grid carbon intensity.  Instead of maintaining a constant production (like the previous scenario), the electrolyzer adjusts its production rate based on the real-time life cycle carbon intensity of the grid's incoming electricity. More specifically, and as shwon in Fig. \ref{fig:Production_Rule_H2}, a predefined production rule is applied so that the carbon intensity (on the x-axis) is mapped to a production rate on the y-axis.  In effect, the rule defines discrete production levels in increments of 2 kg H$_2$, aligning output with the cleanliness of the grid. If the carbon intensity of electricity is less than or equal to 14.50 kg CO$_2$eq per kg H$_2$ produced, the electrolyzer operates at full capacity, producing 20 kg of hydrogen. For moderate carbon intensity levels, such as those between 16.99 kg and 17.00 kg CO$_2$eq/kg H$_2$, the electrolyzer reduces output to 8 kg. When the carbon intensity exceeds 19 kg CO$_2$eq/kg H$_2$, the electrolyzer is turned off entirely to avoid high-emission operation. This controlled production strategy provides an effective mechanism for aligning hydrogen production with periods of lower grid emissions.  This approach leverages the temporal granularity of the carbon intensity dataset to align hydrogen generation with greener grid conditions; aiming to reduce overall life cycle emissions without considering electricity pricing. This production rule serves as a positive step toward environmentally responsive production and provides a comparative basis for understanding how time-varying production can improve the carbon footprint of hydrogen production systems.

\begin{figure}[H]
\centering
\includegraphics[width=0.9\linewidth]{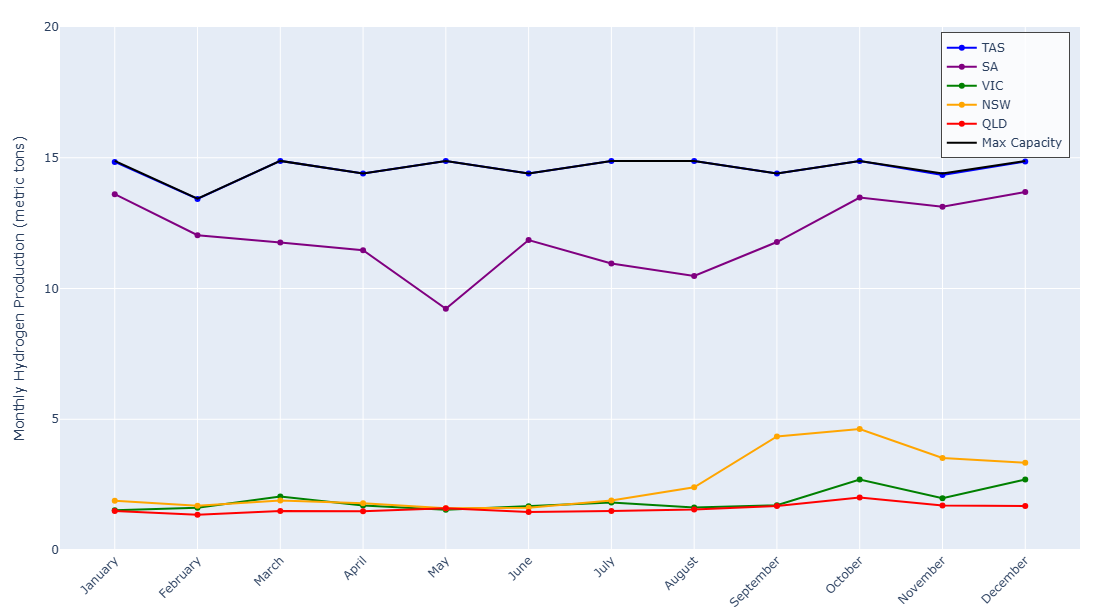}
\caption{Variable Production Scenario 1 in 2023 for: a) New South Wales (orange), b) Queensland (red), c) South Australia (purple), d) Tasmania (blue), and e) Victoria (green). Monthly time series of H\textsubscript{2} production (metric tons).}
\label{fig:H2_Output_All_Months_S2}
\end{figure}

Figure \ref{fig:H2_Output_All_Months_S2} presents the monthly hydrogen production (in metric tons) associated with this variable production scenario.  Of all five states studied, only Tasmania can maintain its maximum production capacity.  Similarly, South Australia achieves a relatively high total hydrogen output due to its frequent periods of high renewable energy penetration.  In contrast, New South Wales, Victoria, and Queensland see lower annual production where grid carbon intensity remains elevated for longer periods of time.

\begin{figure}[H]
\centering
\includegraphics[width=0.9\linewidth]{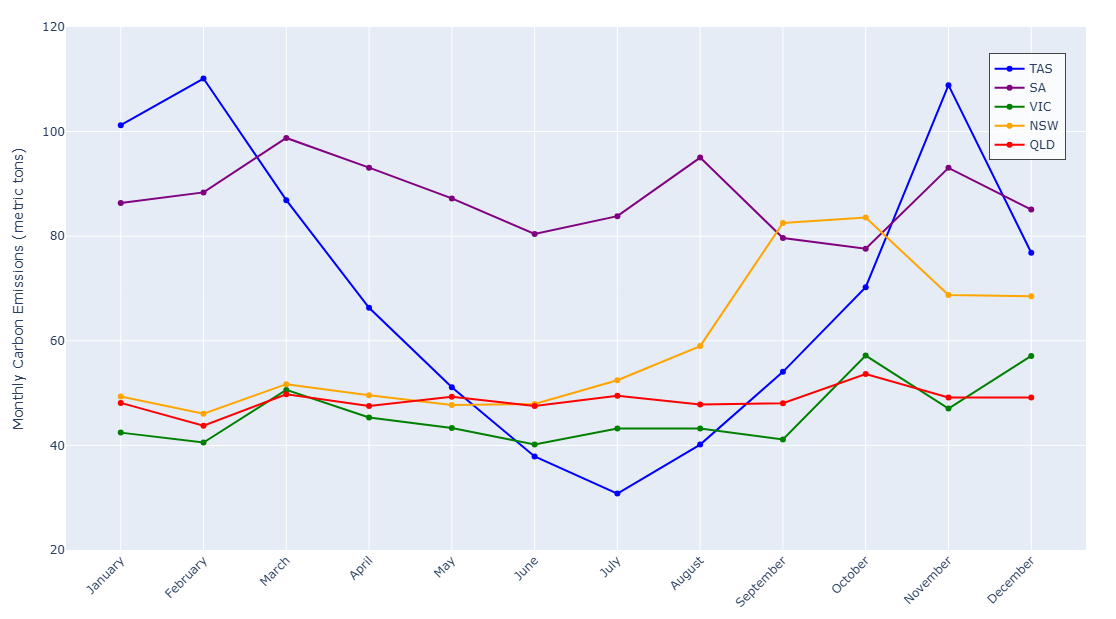}
\caption{Variable Production Scenario 1 in 2023 for a.) New South Wales (orange), b.) Queensland (red), c.) South Australia (purple), d.) Tasmania (blue), e.) Victoria (green) in 2023.  Monthly time series of per unit H$_2$ cost in $\$/kg$.}
\label{fig:CO2_output_all_months_S2}
\end{figure}

Figure \ref{fig:CO2_output_all_months_S2} shows the monthly life cycle carbon emissions (in metric tons CO$_2$eq) associated with the variable hydrogen production shown in Figure \ref{fig:H2_Output_All_Months_S2}. Compared to the baseline scenario shown in Fig. \ref{fig:CO2_output_all_months_S1}, the emissions are significantly reduced across all states, particularly in Tasmania and South Australia. These reductions are a direct result of restricting hydrogen production to hours when the grid carbon intensity falls below predetermined thresholds. The figure confirms that this operational strategy is effective in minimizing the environmental footprint of electrolytic hydrogen without requiring changes to infrastructure or supply sources. It provides clear evidence of the emissions-saving potential of time-sensitive, grid-aware production scheduling.

\begin{figure}[H]
\centering
\includegraphics[width=0.9\linewidth]{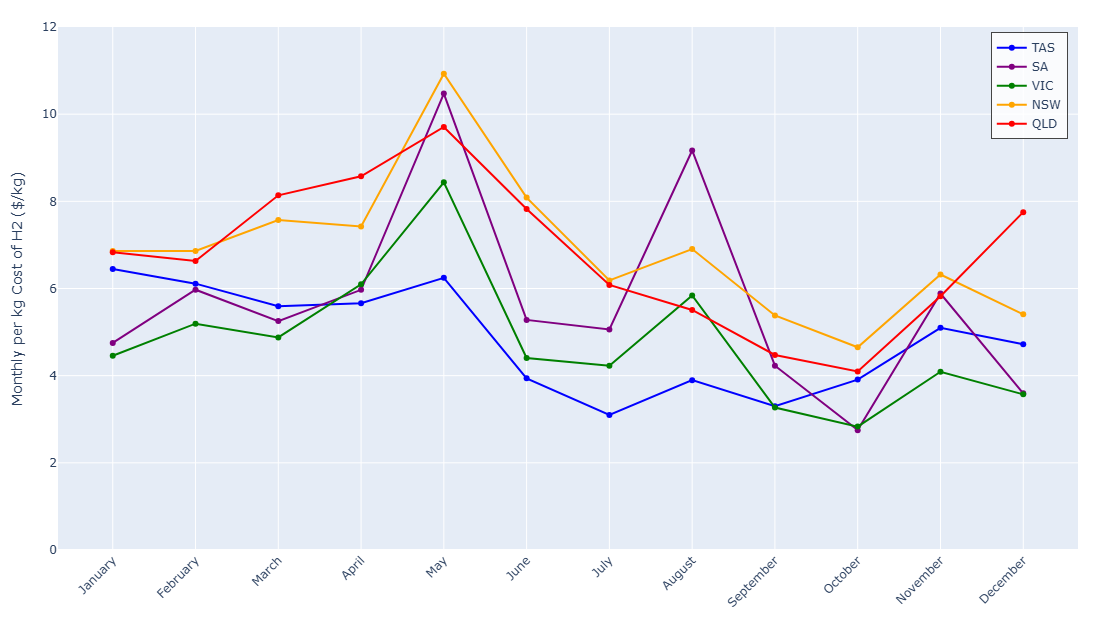}
\caption{Variable Production Scenario 1 in 2023 for a.) New South Wales (orange), b.) Queensland (red), c.) South Australia (purple), d.) Tasmania (blue), e.) Victoria (green) in 2023.  Monthly time series of life cycle carbon emissions in metric tons.}
\label{fig:S2_Per_kg_h2_Cost}
\end{figure}

Figure \ref{fig:S2_Per_kg_h2_Cost} shows the monthly per unit cost of hydrogen production in (\$AUD/kg H$_2$) for each state in the variable production scenario.  Although this scenario was not explicitly designed to take advantage of low electricity prices, the hydrogen production in low-carbon hours coincidentally overlaps with periods of higher renewable generation and lower electricity prices.  As a result, states like Tasmania exhibit lower per unit Hydrogen cost due to frequent low-price hours, while states with less renewable penetration, such as New South Wales, Victoria, and Queensland, experience higher production costs due to its grid's high dependency on non-renewable energy.  

\subsection{Smart Production Scenario: “Exploit Tax Incentives”}\label{Sec:SmartProduction}
This third scenario intensifies the dynamic operational strategy of the previous variable production scenario.  More specifically, hydrogen production is restricted to hours when the life cycle carbon intensity (LCA CO$_2$eq) of electricity falls below a predefined threshold of 0.6 kg CO$_2$ per kg H$_2$ that qualifies for hydrogen tax credits.  The objective is to receive maximum financial incentives, such as those outlined in Section 45V of the U.S. Inflation Reduction Act or similar mechanisms proposed in the Australian policy context (even if it results in lower overall hydrogen production).   By aligning electrolyzer operation with both real-time grid carbon intensity and regulatory eligibility criteria, this scenario aims to reduce emissions while also lowering the effective cost of hydrogen production. This scenario represents a scenario where environmental performance and economic viability are jointly considered; offering a scalable and policy-compliant hydrogen generation.

\begin{figure}[H]
\centering
\includegraphics[width=0.9\linewidth]{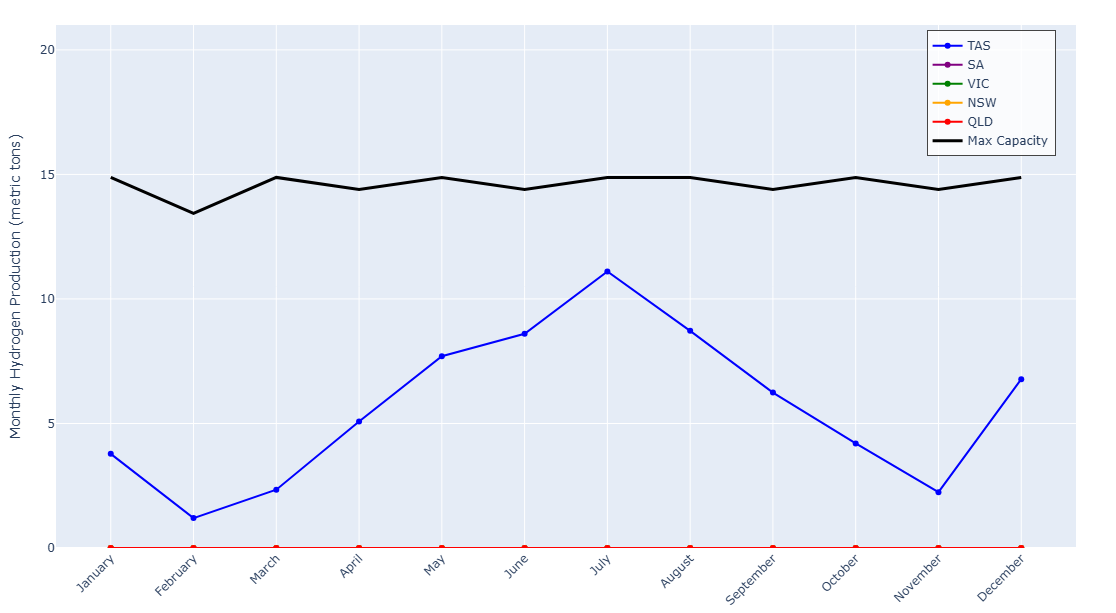}
\caption{Smart Production Scenario 2 Relative Monthly H$_2$ production for the Five States of Australia. Electrolyzer capped at a carbon intensity of 0.6 kg CO$_2$eq per kg of hydrogen Production (Max Capacity 20kg H$_2$/hr) Capacity Over 1 Year Span.}
\label{S3_H2_Output_All_Months}
\end{figure}

Figure \ref{S3_H2_Output_All_Months} shows that the monthly hydrogen production (in metric tons) is substantially lower, as the electrolyzer operates only during hours that meet the strict emissions threshold of 0.6 kg CO$_2$ per kg H$_2$. This limited operating window significantly constrains hydrogen output, where fossil fuel-based electricity often dominates the grid mix. Hence, no hydrogen is produced in the states of South Australia, Victoria, New South Wales, and Queensland. In contrast, Tasmania becomes the only state with sufficiently clean electricity to have any hydrogen production.  

\begin{figure}[H]
\centering
\includegraphics[width=0.9\linewidth]{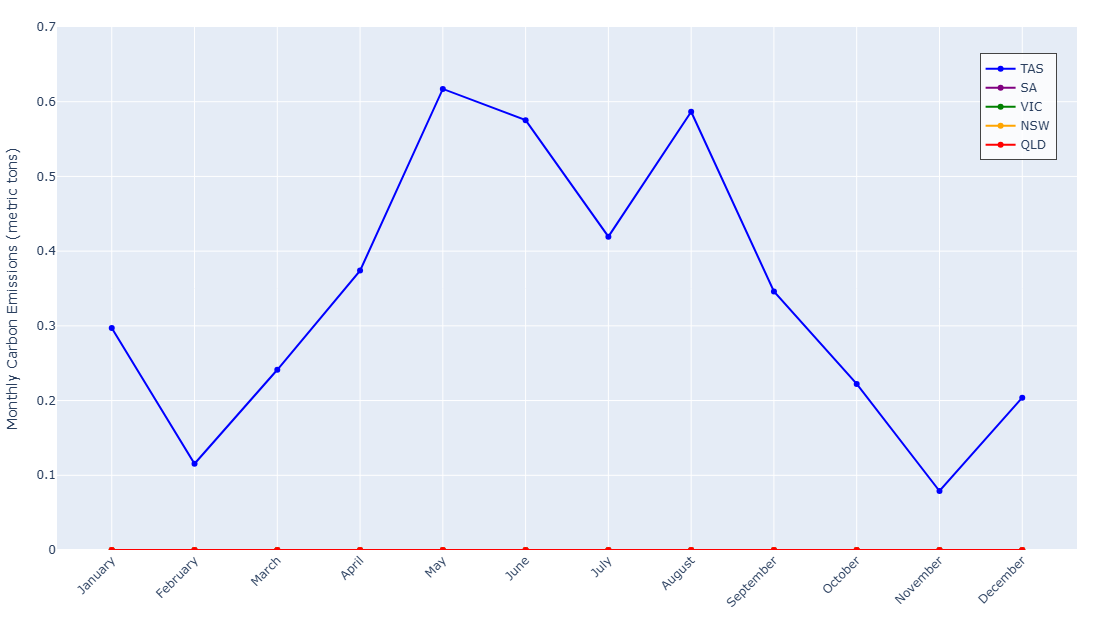}
\caption{Smart Production Scenario 2 Relative Monthly H$_2$ Carbon Emissions for the Five States of Australia. Electrolyzer capped at a carbon intensity of 0.6 kg CO$_2$eq per kg of hydrogen Production (Max Capacity 20kg H$_2$/hr) Capacity Over 1 Year Span.}
\label{S3_CO2_output_all_months}
\end{figure}

Figure \ref{S3_CO2_output_all_months} displays the monthly life cycle carbon emissions (in metric tons of CO$_2$ equivalent) for the smart production scenario.  Emissions are recorded only in Tasmania, the sole state meeting the 0.6 kg CO$_2$/kg H$_2$ threshold, as shown in Figure \ref{S3_H2_Output_All_Months}. All other states report zero emissions due to the absence of hydrogen production.

\begin{figure}[H]
\centering
\includegraphics[width=0.9\linewidth]{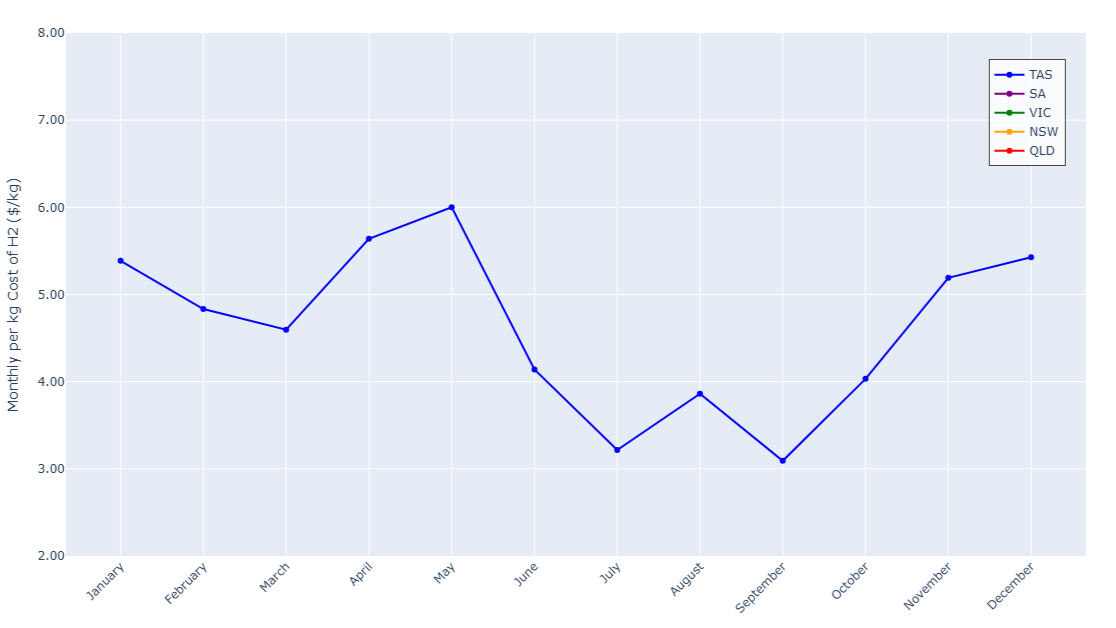}
\caption{Smart Production Scenario 2 Relative Monthly H$_2$ per kg Cost for the Five States of Australia. Electrolyzer capped at a carbon intensity of 0.6 kg CO$_2$eq per kg of hydrogen Production (Max Capacity 20kg H$_2$/hr) Capacity Over 1 Year Span.}
\label{S3_Per_kg_h2_Cost}
\end{figure}

Figure \ref{S3_Per_kg_h2_Cost} shows the monthly average hydrogen production cost (in AUD/kg H$_2$) in the Smart Production Scenario (without policy incentives).  As only Tasmania meets the strict carbon intensity threshold, cost data is available solely for this state. Despite limited production hours, the hydrogen produced benefits from low electricity prices during clean grid periods, as can be seen in July and October, resulting in relatively low average costs of approximately AUD 3 per kg H$_2$. No cost data appears for other states due to zero production under this scenario.  

\begin{figure}[H]
\centering
\includegraphics[width=0.8\linewidth]{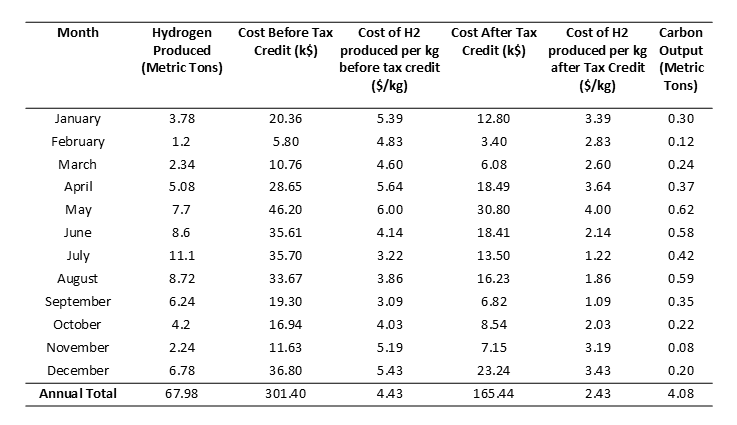}
\caption{Monthly Breakdown of Hydrogen Production, Carbon Output, Costs, and Tax Credit Earnings for Tasmania under Scenario 3.}
\label{table_scenario_3_outcome}
\end{figure}

Figure \ref{table_scenario_3_outcome} presents the monthly breakdown of hydrogen production, carbon emissions, production costs, and associated tax credit earnings for Tasmania under the Smart Production Scenario \ref{Sec:SmartProduction}. As the only state that meets the stringent carbon intensity threshold of 0.6 kg CO$_2$eq per kg H$_2$, Tasmania exemplifies how policy-aligned hydrogen production can operate effectively under clean energy constraints. The figure reveals that hydrogen generation is concentrated in months such as July and October, when favorable grid conditions coincide with tax credit eligibility. This temporal alignment results in reduced production costs and maximized financial returns despite limited operating hours. Tax credit earnings reflect this trend, underscoring the economic benefits of synchronizing electrolyzer operation with emissions-based policy incentives. Overall, the results demonstrate that under strict regulatory conditions, clean hydrogen production remains technically and economically feasible in regions with consistently low-carbon electricity.

\begin{figure}[H]
\centering
\includegraphics[width=1\linewidth]{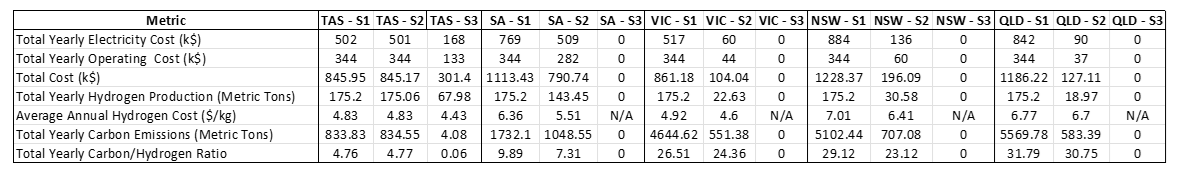}
\caption{Comparison of yearly analysis across five Australian states (TAS, SA, VIC, NSW, QLD) under three operational scenarios (S1: Baseline Scenario, S2: Variable Production Scenario 1, S3: Smart Production Scenario "Exploit Tax Incentives"). Metrics include total yearly electricity cost, operating cost, total cost, hydrogen production, hydrogen cost, carbon emissions, and carbon/hydrogen ratios, highlighting the direct coupling of operating costs with electricity usage in the absence of optimization.}
\label{fig:Comparison_S1_S2_S3_Table}
\end{figure}

Figure \ref{fig:Comparison_S1_S2_S3_Table} provides a comprehensive comparative overview of total yearly electricity costs, operating costs, carbon emissions, hydrogen output, and carbon-to-hydrogen ratios across all five Australian states under the three defined production scenarios. In Scenario 1, the Baseline Scenario: Fixed Hydrogen Production (\ref{Sec:BaselineScenario}), hydrogen production remains fixed at 20 kg/hr, which is 175.2 metric tons across all states, regardless of temporal variations in grid carbon intensity or electricity price. This approach results in the highest carbon emissions and carbon-to-hydrogen ratios, particularly in states with fossil fuel-heavy grids such as Queensland (31.79) and New South Wales (29.12). Operating costs mirror electricity expenditures due to the lack of any intelligent optimization. Scenario 2 Variable Production Scenario 1: Produces H$_2$ When the Grid is Green (\ref{Sec:AusCharacteristics}), demonstrates a marked improvement in emissions performance by reducing hydrogen output in response to high-carbon periods, thereby lowering both emissions and carbon intensity ratios. For example, Victoria's ratio drops from 26.51 to 24.36 while maintaining some level of production. However, the gains come at the cost of reduced hydrogen volumes, especially in states where clean energy availability is less frequent.

Smart Production Scenario 2: “Exploit Tax Incentives” (\ref{Sec:SmartProduction}) imposes the most stringent operational constraint, producing hydrogen only when the life cycle carbon intensity is below 0.6 kg CO$_2$ per kg H$_2$, aligned with Australia's proposed clean hydrogen certification standards and tax incentive eligibility. As a result, production ceases entirely in South Australia, Victoria, New South Wales, and Queensland, leading to zero emissions and operating costs in these regions. Only Tasmania, with its predominantly renewable grid, continues to generate hydrogen (67.98 metric tons) under this scenario at a drastically lower carbon/hydrogen ratio of 0.06. This stark contrast illustrates the policy-aligned potential for near-zero-carbon hydrogen when production is tightly coupled with both environmental and economic signals. The table highlights how the potential for clean hydrogen production varies across Australian states and shows that grid carbon intensity plays a key role in making hydrogen production scalable, affordable, and eligible for policy incentives.

\section{Conclusion and Future Work}\label{Sec:Conclusion}
This study highlights that optimizing hydrogen production by aligning it with periods of low-carbon electricity availability can significantly reduce both emissions and costs, particularly in states with substantial renewable energy contributions like South Australia and New South Wales. In contrast, coal-dependent regions such as Victoria and Queensland face greater challenges, with more limited reductions in carbon intensity and cost savings. The integration of real-time data on carbon intensity and electricity prices proves to be an effective strategy for maximizing production efficiency, though further developments are required to unlock its potential fully. Key areas for improvement include increasing renewable energy integration, implementing storage solutions to manage renewable variability, and establishing supportive policy frameworks like hydrogen tax credits and carbon pricing. Further exploration of these aspects, along with applying this optimization approach in other regions with diverse energy grids, will enhance the scalability and impact of hydrogen as a sustainable energy solution.

\bibliographystyle{IEEEtran}
\bibliography{LIINESLibrary,LIINESPublications,LCA}
\end{document}